%% file: w3h2o_pol.tex
\documentclass[structabstract]{aa}  % version 7
\usepackage[varg]{txfonts}
\usepackage{amssymb}
\usepackage{natbib}
\bibpunct{(}{)}{;}{a}{}{,}
\usepackage{amsmath} 
\usepackage{color}
\newcommand{\pas}{$\rlap{.}^{\prime\prime}$}
\newcommand{\dg}{$^{\circ}$}
\newcommand{\kms}{km~s$^{-1}$}
\newcommand{\cmc}{cm$^{-3}$}

\def\wat {H$_2$O}

\def\dg{$^{\circ}$}
\def\kms{km\,s$^{-1}$}

\def\jyb{Jy\,beam$^{-1}$}
\def\mjyb{mJy\,beam$^{-1}$}

\def\ls {\hbox{L$_{\odot}$}} 
\def\d {$^{\circ}$}

\def\tbo {$T_{\rm{b}}\Delta\Omega$}

\def\dvi {$\Delta V_{\rm{i}}$}
\def\dvl {$\Delta v_{L}$}

\def\pl {$P_{\rm{l}}$}
\def\w3{W3(H$_2$O)}
\def\vlsr{V$_{\rm LSR}$}

\begin{document}

\title{Measuring Magnetic Fields from Water Masers Associated with the Synchrotron Protostellar Jet in W3(H$_2$O)}

\author{C. Goddi  \inst{1,2}  
\and G. Surcis   \inst{3,4}  
\and L. Moscadelli  \inst{5}
\and H. Imai \inst{6}
\and W. H. T. Vlemmings  \inst{7}
\and  H. J. van Langevelde \inst{3,8}
\and A. Sanna  \inst{9}
}
\institute{Department of Astrophysics/IMAPP, Radboud University, PO Box 9010, NL-6500 GL Nijmegen, the Netherlands
\and  ALLEGRO/Leiden Observatory, Leiden University, PO Box 9513, NL-2300 RA Leiden, the Netherlands 
\and Joint Institute for VLBI in Europe, Postbus 2, NL-79990 AA Dwingeloo, The Netherlands
\and INAF - Osservatorio Astronomico di Cagliari
Via della Scienza 5 - I-09047 Selargius, Italy
\and INAF, Osservatorio Astrofisico di Arcetri, Largo E. Fermi 5, I-50125 Firenze, Italy
\and Department of Physics and Astronomy, Kagoshima University, 1-21-35 Korimoto, Kagoshima 890-0065, Japan 
\and Chalmers University of Technology, Onsala Space Observatory, 43992 Onsala, Sweden
\and Sterrewacht Leiden, Leiden University, PO Box 9513, NL-2300 RA Leiden, the Netherlands
\and Max-Planck-Institut  f\"{u}r Radioastronomie, Auf dem H\"{u}gel 69, D-53121 Bonn, Germany
 }

\date{Received ; accepted}

\abstract
%context heading (optional)
{Magnetic fields are invoked to launch, drive, and shape  jets in both low- and high-mass protostars, but observational data on the  spatial scales required to assess their role in the protostellar mass-loss process is still scarce.}
{The Turner-Welch (TW) Object in the W3(OH) high-mass star forming complex drives a synchrotron jet, which is quite exceptional for a high-mass protostar, 
and is associated with a strongly polarized \wat\,maser source, \w3, 
 making it an optimal target to investigate the role of magnetic fields on the innermost scales  of protostellar disk-jet systems. 
}
{We report full polarimetric VLBA observations of H$_{2}$O masers towards  \w3.  
Their linearly  polarized emission provides clues on the orientation of the local magnetic field (on the plane of the sky), while the measurement of the Zeeman splitting provides its strength (along the line-of-sight). The linear scales probed by \wat\,masers are tens to hundreds of AU (at the \w3\ distance, $\sim  2$~kpc), 
inaccessible to other star formation tracers. }
{We identified a total of 148 individual maser features and we measured their physical properties. 
Out of 148, we measured linear polarization in 34  features, with a fractional percentage varying in the range 0.9\%--42\%, making \w3\,the highest polarized \wat\,maser source observed with VLBI known in the Galaxy. 
The \wat\,masers trace a bipolar, biconical outflow at the center of the synchrotron jet. 
Although  on scales of a few thousand AU the magnetic field inferred from the masers is on average orientated along the flow axis, on smaller scales (10s to 100s of AU), we have revealed a misalignment between the magnetic field and the velocity vectors, which arises from the compression of the field component along  the shock front.  
We also detected circularly polarized emission toward 10 maser features, with a fractional percentage varying in the range 0.2--1.6\%. 
In the  gas shocked by the synchrotron jet, we estimate a total field strength  in the range $\sim$100-300~mG  (at densities of $10^9$~\cmc). We conclude that fields of this order of magnitude are expected if the  observed polarized water masers  emerge behind magnetically supported shocks which, propagating in the \w3\ hot core (with an initial  density of order of $10^7~\rm{cm^{-3}}$), compress and enhance the field component perpendicular to the shock velocity (with an initial field strength of a few mG). 
We constrain  the magnetic field strength in the pre-shock circumstellar gas (which is dominated by the component parallel to the flow motion) to at least 10--20~mG  (at densities of $10^7$~\cmc),  consistent with previous estimates from a synchrotron jet model and dust polarization measurements.
}
{
In \w3, the magnetic field would evolve from having a dominant  component parallel to the outflow velocity in the pre-shock gas, with field strengths of the order of a few tens of mG,  to being mainly dominated by  the perpendicular component of order of a few hundred of mG in the post-shock gas where the \wat\,masers are excited. 
The general implication is that in the  undisturbed (i.e. not-shocked) circumstellar gas, the flow velocities would follow closely the magnetic field lines, while in the shocked gas the magnetic field would be re-configured to be parallel to the shock front.  
}


\titlerunning{The magnetic field in a protostellar synchrotron jet.}
\authorrunning{Goddi et al.}

\maketitle
%________________________________________________________________

%================================================================================================================================
\section{Introduction}

Protostellar jets from both low-mass and high-mass Young Stellar Objects (YSOs) are known to be ionized, primarily through shocks, during early stages of their evolution \citep{Anglada1998,Hofner2011}. 
In the case of high-mass YSOs, the H-ionizing luminosity increases dramatically when the YSO reaches the main sequence, resulting in prominent ionized winds and/or compact HII regions \citep{Hoare2007}. 
Therefore,  thermal (bremsstrahlung) emission is often associated with (high-mass) YSOs, 
and observations of radio continuum as well as  recombination lines can be naturally used to study the ionized component of (high-mass) protostar outflows. 

Besides (thermal) free-free emission due to shock-induced ionisation, low-mass YSOs like  T Tauri stars are known to emit also  (non-thermal) gyrosynchrotron emission from mildly relativistic electrons gyrating in magnetic fields \citep{Andre1996}. 
Non-thermal emission in radio jets  has been also observed in some massive protostars, e.g. \w3\, \citep{Reid1995}, HH 80-81 \citep{Carrasco2010}, IRAS 18182-1433 \citep{Hofner2011}, and G16.59-0.05 \citep{Moscadelli2013}. 
These findings have been interpreted in terms of synchrotron emission from relativistic electrons accelerated in strong shocks, but overall synchrotron jets appear to be quite rare. 
However,  a recent study by \citet{Moscadelli2016}  suggests that non-thermal continuum emission could be common in high-mass protostellar jets, after all. Studying the magnetic field in  synchrotron jets provides a great opportunity to understand its role in the protostellar mass-loss process and, more in general, in high-mass star formation (HMSF). 

The Turner-Welch (TW) Object \citep{TurnerWelch84} is a luminous hot-core ($L \sim 10^4$~\ls)
about 6\arcsec\, or $ 10^{4}$ AU east of the archetypal ultracompact (UC) HII region W3(OH), at a distance of 2.04~kpc \citep{Hachi06}.  
With respect to the neighbouring W3(OH), it shows weak OH masers and no UC-HII region, indicating an earlier evolutionary stage. 
The dense molecular clump  is forming a protobinary (or a multiple) system, as inferred from the structure in the mm dust emission \citep{Wyrowski1999,Zapata2011}. 
Remarkably, the most massive member of the multiple  (labelled "A"; \citealt{Wyrowski1999}) is driving a synchrotron jet  \citep{Reid1995,Wilner1999},  along the east-west (E-W) direction, 
which is also the direction along which the multiple is forming. 
Besides the synchrotron jet, the TW-object powers  strong H$_2$O masers, therefore the core is also indicated as \w3. 
Proper motions measurements of these water masers revealed a bipolar outflow centered at the synchrotron jet, expanding at about 20~\kms\, \citep{Alcolea1993,Hachi06}. 

One interesting property of molecular masers is that, besides being superb kinematic probes,  they can be polarized. 
With a detailed theory of maser polarization propagation, polarimetric observations can yield 
the strength of the magnetic field along the   line-of-sight (l.o.s.)  from circular polarization 
(Zeeman effect) and the 2D (or even the 3D) field structure from linear polarization (see \citealt{Vlemmings2012}, for a review). 
 And indeed a number of water maser sources have been studied with polarization observations, and all showed that  the linear polarization vectors of individual water maser features are generally aligned with each other,  
indicating that the measured polarization truly probes the magnetic field in the maser region   
\citep{Imai2003,Vlemmings2006b,Surcis2011a,Surcis2011b,Surcis14}. 

In this Paper, we present full polarimetric observations of \wat\,masers in \w3, with the main goal of measuring for the first time the magnetic field strength and structure along the synchrotron jet  driven by a high-mass protostar. 
The general question we want to address is whether there is a coupling between  the gas motion and the magnetic field, i.e. if the gas motion locally follows the magnetic field or viceversa. In order to address this question, we need to probe  the small scales where jets and winds are expected to be launched and collimated by massive protostars (a few tens to hundreds of AU; see e.g. \citealt{Matthews2010,Greenhill2013}). This requires very long baseline interferometry (VLBI) measurements.

%================================================================================================================================
\section{Observations and data reduction}
\label{obssect}
We observed the massive star-forming region W3(OH) in the 6$\rm{_{16}}$-5$\rm{_{23}}$ transition of \wat ~(rest frequency: 
22.23508~GHz) with the NRAO\footnote{The National Radio Astronomy Observatory (NRAO) is a facility of the National Science Foundation 
operated under cooperative agreement by Associated Universities, Inc.} Very Long Baseline Array (VLBA) on August 18$\rm{^{th}}$ 2006. 
The observations were conducted 
in full polarization spectral mode using one single baseband filter of 2 MHz, which covered a total velocity range of
$\approx27$~\kms. We performed three correlation passes in total. 
The first one had 128 channels with a spectral resolution of 15.6~kHz (0.2~\kms) and included all 4 polarization combinations: RR, LL, RL,  LR. 
This pass  allowed to produce images in all Stokes parameters and 
was used in the analysis to measure the linear polarization of individual maser features. 
The other two passes had higher spectral resolution (1.95~kHz or 0.03~\kms) and 1024 channels, 
and contained only the circular polarization combinations: RR and LL, respectively. 
These extra two passes were required to measure Zeeman splitting of individual \wat~maser lines. 
Including the overheads, the total observation time was 12~h.

The data were calibrated and imaged using the Astronomical Image Processing Software package (AIPS).  
The calibration of the bandpass, delay, rate, and phase was performed on the calibrators 0420-014 and DA193. 
The fringe-fitting and the self-calibration were performed on the brightest maser feature at V$_{\rm LSR}$=--48.6~\kms\ (Table~\ref{features}),  on the dataset with 0.2~\kms\ velocity resolution. 
In order to calibrate the polarization, we first removed instrumental effects, by estimating the D-terms for each antenna.  
For this purpose, we used 0420-014, which is a linearly polarized source \citep[2\%;][]{Marscher02}, 
covering a large variation ($>$ 90\dg) of the parallactic angle during the observations. 
The solutions from fringe-fitting and D-terms were then transferred from the moderate to the high spectral resolution dataset. 
We did not correct for the R-L delay offset, which affects  the polarization angle  but in a negligible way with respect to its uncertainty.

Imaging and deconvolution of the two fully calibrated data sets were
performed within the AIPS task IMAGR, using Briggs weighting  with ${\cal R}$=0 and a pixel size of 0\pas0002, 
which resulted in a beam-size of  0\pas00097 $\times$ 0\pas00074.  
The phase-center position was: $\alpha(J2000) = 02^h 27^m 04\rlap{.}^s836$,   $\delta(J2000) = +61^{\circ} 52' 24.61"$, 
corresponding to the TW-object. 
Since the water maser distribution extends E-W across more than 3\arcsec\,\citep{Hachi06}, 
we simultaneously imaged three fields of 8192$\times$8192 pixels or 1\pas6$\times$1\pas6 each around the  TW object,  
sufficient to cover the total area with known water maser emission.  
A fourth field was imaged approximately 7\arcsec\,to the West of the  TW-object, at the position of the HII region W3(OH), 
but no water maser emission was detected. 
For each field, we created  data cubes  of Stokes 
{\it I} ($rms=7.3$~\mjyb), {\it Q} ($rms=6.5$~\mjyb) and {\it U} ($rms=6.5$~\mjyb)  for the modest spectral resolution dataset, 
and Stokes {\it I} and {\it V} ($rms=10$~\mjyb) for the high spectral resolution dataset. 
The {\it Q} and {\it U} cubes were combined
 to produce cubes of polarized intensity ($POLI=\sqrt{Q^{2}+U^{2}}$) and polarization angle ($POLA=1/2\times~atan(U/Q)$). 
 We calibrated the linear polarization angles of individual \wat~masers by rotating the linear polarization angle 
measured for 0420-014 from our VLBA data to the one measured  
in a POLCAL VLA observation\footnote{http://www.aoc.nrao.edu/$\sim$smyers/calibration/} 
carried out 43 days before our VLBA observations:  
$\chi_{\rm{0420-014}}=+66$\d$\!\!.91$. 
The formal errors on the $POLA$ values for individual spectral channels are due to thermal noise: 
$\sigma_{\rm POLA}=0.5 ~\sigma_{P}/P \times 180^{\circ}/\pi$  \citep{war74}, 
where $P$ and $\sigma_{P}$ are the polarization intensity and corresponding rms error, respectively. 
The I and
V cubes at high spectral resolution were used to evaluate the magnetic
field strength along the line of sight, which is proportional to the circular 
polarization fraction $P_{\rm{V}}=(V_{\rm{max}}-V_{\rm{min}})/I_{\rm{max}}$.
 The analysis of the maser polarization data is described in Appendix~\ref{analysis}. 

We did not observe in phase-referencing mode, therefore during self-calibration we lose the information on the absolute position. The positions of the identified \wat\,masers are relative to the strongest maser feature used for self-calibration at --48.6~\kms. 
In order to estimate the absolute position in our maps, 
we used the astrometric measurements by \citet{Hachi06} (the procedure is described in Appendix~\ref{astro}).

%--------------------------------------------------------------------------------------
\begin{figure*}
\centering
\includegraphics[angle=-90,width= \textwidth]{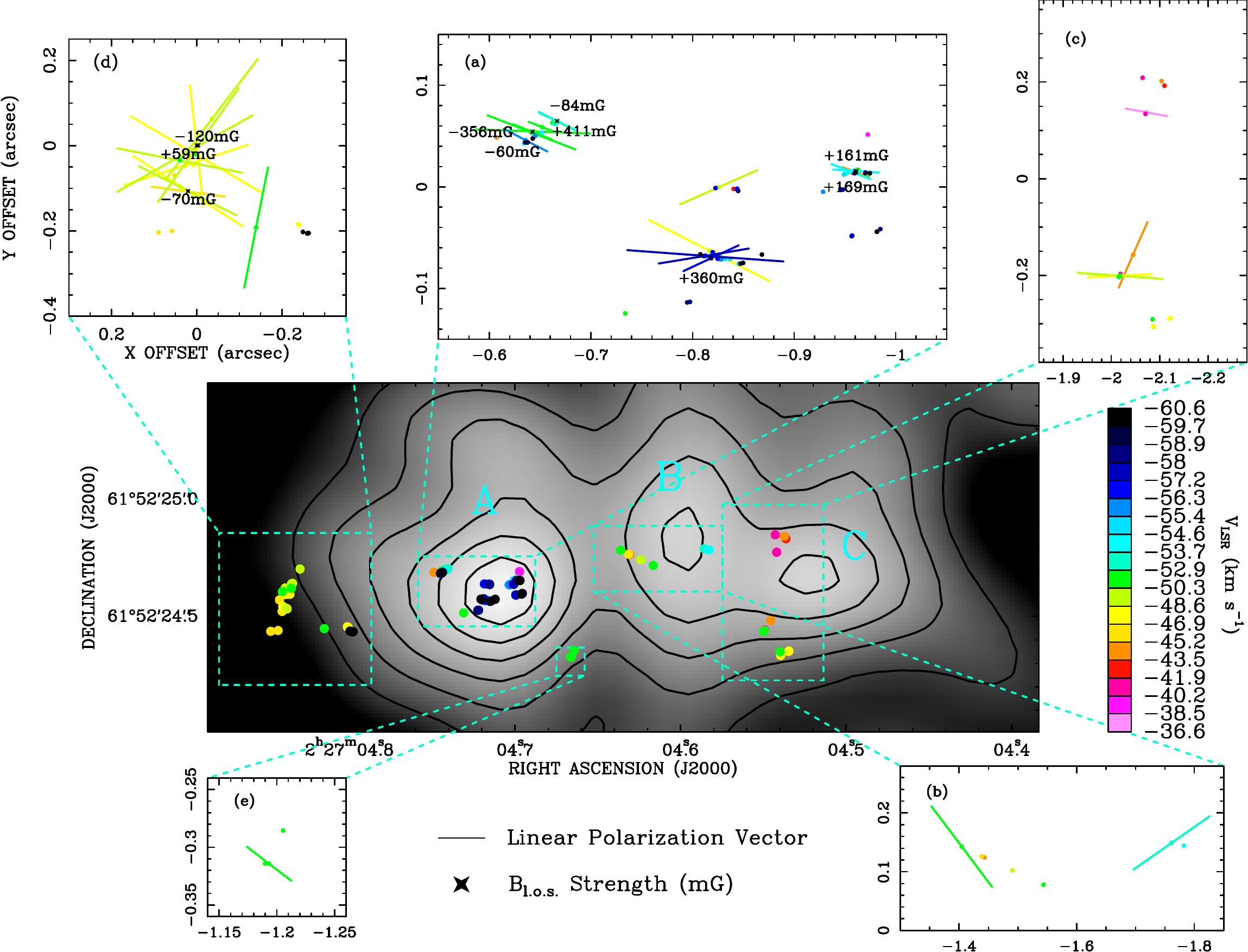}
\caption{Overlay of the water masers detected with the VLBA in \w3\, onto the 1.4~mm continuum emission mapped with the PdBI by \citet{Wyrowski1999}  (gray scale and black contours). 
 The circles show positions of the \wat\,masers, while the colors denote their l.o.s.  velocity in \kms\,(color scale on the right-hand side). 
The three 1.4 mm continuum peaks identified by  \cite{Wyrowski1999} are labelled \ "A", "B",  and "C", from east to west. Contour levels correspond to steps of  24 mJy beam$^{-1}$ (starting from 47 mJy beam$^{-1}$).  
The insets show the linear polarization vectors of individual maser features in different clusters (from "a" to "e"),  where the length of the line segments scales logarithmically with the polarization fraction  (in the range $P_{\rm{l}}=0.9\%-42\%$).  
We also report the  magnetic field strengths (in mG) along the l.o.s. ($B_{\rm l.o.s.}$) in the maser features for which we measured the Zeeman splitting. The positions are relative to the reference maser feature  used for data self-calibration (ID 018 in Table~\ref{features}). 
}
\label{pol_map}
\end{figure*}
%%--------------------------------------------------------------------------------------

%--------------------------------------------------------------------------------------
\begin{figure*}
\centering
\includegraphics[angle=-90,width= \textwidth]{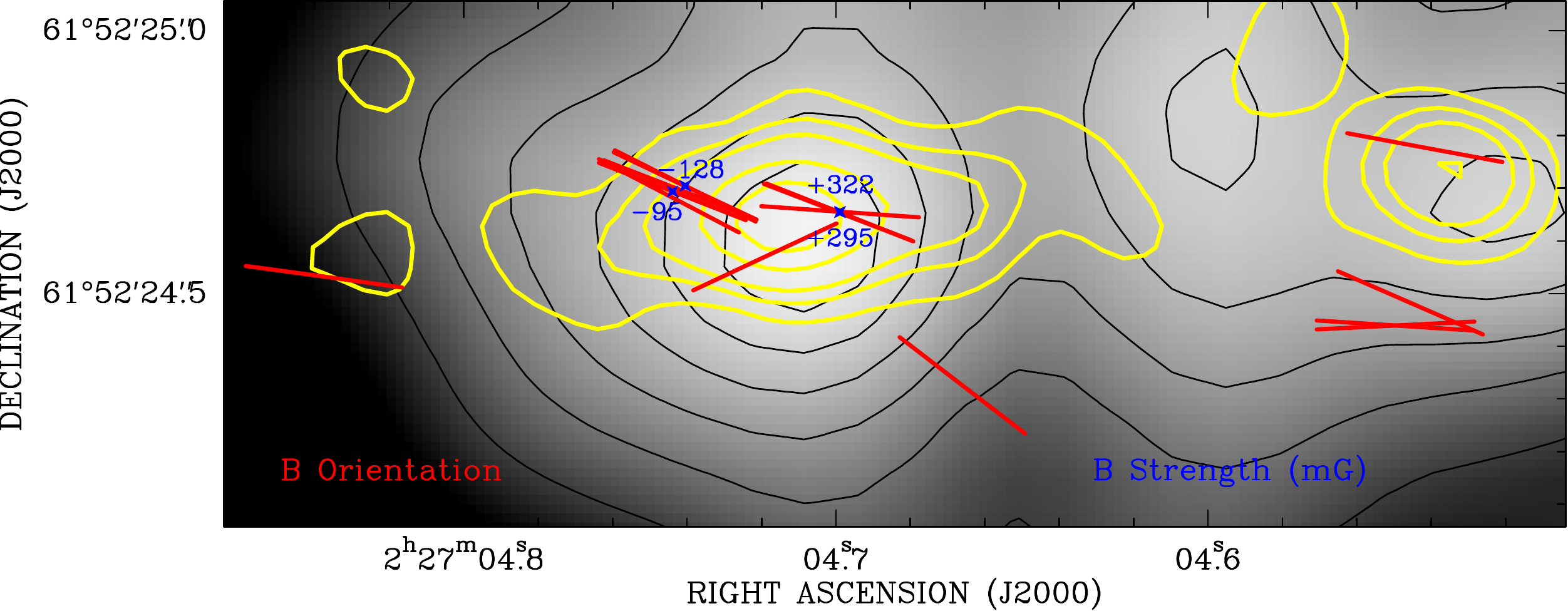}
\caption{Magnetic field orientation (in the plane of the sky) for 17 individual masers with $P_l<5$\% (red segment) and strength for 4 (non-saturated) masers for which the Zeeman splitting was measured.  
The 8.4 GHz emission imaged with the VLA (beamsize$\sim$0\pas2) by \citet{Wilner1999}  (yellow contours: corresponding to  0.02, 0.06, 0.1, 0.2, 0.3 mJy beam$^{-1}$) is overploted onto the 1.4~mm continuum emission mapped with the PdBI  (beamsize$\sim$0\pas5)  by \citet{Wyrowski1999} 
 (gray scale and black contours: same as in Fig.~\ref{pol_map}). 
The radio continuum shows a main central component,  the synchrotron jet, and two (western and eastern) secondary components (see \S~\ref{B_dyn} for an interpretation).  
}\label{B_map}
\end{figure*}
%--------------------------------------------------------------------------------------

%========================================================================================================
\section{Results}
\label{res}
We used the VLBA  at 22~GHz to perform full polarization observations of water masers toward  \w3.  
We identified a total of 148 individual maser features and we measured their physical properties, including positions, flux densities, l.o.s. velocities (\vlsr), and (when polarized) their fraction of linear and circular polarizations, as well as the corresponding linear polarization angles and  magnetic field strengths  along the  l.o.s.
(these parameters are listed in Table~\ref{features}). 
The methodology adopted to derive the physical properties of individual maser features is described in Appendix~\ref{analysis}. 

The identified maser features have peak flux densities from 50 mJy to about 2000 Jy, and \vlsr\, spanning from --60.6 to --39.7~\kms.  
Figure~\ref{pol_map} shows positions and l.o.s. velocities of the 22 GHz water maser features overplotted  on the contour map of the dust continuum emission at 1.4~mm, imaged with the PdBI by \citet{Wyrowski1999}. 
The masers are distributed over an area of 2\pas5 $\times$ 0\pas5 along E-W, 
and are clustered in five main groups, which we label with letters from "a" to "e". 
The l.o.s. velocities do not show any particularly ordered distribution, consistent with a flow in the plane of the sky \citep{Reid1995}. 
The 1.4~mm continuum is elongated E-W, across approximately 3\arcsec, and comprises three dusty components, labelled "A", "B", and "C" from east to west \citep{Wyrowski1999}.  
The masers are loosely associated with these three peaks of dust emission with the majority of  \wat\ masers detected  towards the strongest dust continuum peak ("A") and to the east from it, at the eastern "tip" of the elongated dusty core, where also the highest intensity maser features  are observed. 

 Besides positions and l.o.s. velocities, we also measured the position angle of the linear polarization vectors ($\chi$) and the Zeeman splitting for single  maser features. We describe these measurements in the next two subsections. 
 
%-------------------------------------------------------------
\subsection{Linear polarization: Magnetic Field Orientation}
\label{lin_pol}
%-----------------------------------------------------
Out of  the 148 features detected, we measured linear polarization in 34 maser features, with a fractional percentage varying in the range $P_{\rm{l}}=0.9\%-42\%$ (see Table~\ref{features} and Appendix~\ref{analysis}). 
In Figure~\ref{pol_map},  line segments  indicate the linear polarization vectors of the maser features, whose  length scales logarithmically with the polarization fraction, $P_l$.
Remarkably, 13 features have $P_{\rm{l}} > 10$\%, with 4 having $P_{\rm{l}} > 35$\%, making \w3\,the highest polarized \wat\,maser source  on VLBI scales
known in the Galaxy\footnote{Before this work, the highest polarization fraction of water masers (26\%) measured with VLBI was found in W75N \citep{Surcis14}. Besides VLBI, higher fractions of linear polarization were detected with single-dish measurements (e.g.,  \citealt{Garay1989} registered a 60\% fraction in a \wat\,maser flare in Orion BN-KL)}. 
 This is both a blessing and a curse, because on the one hand 
 such high fractions of polarization are obviously easier to measure (flux-wise), but on the other hand  these masers are most probably saturated, and their polarized signal does not reflect the source magnetic field (see for example the discussion in \citealt{Garay1989} in the case of the \wat\,maser flare in Orion BN-KL). 
In particular, this hampers the exact relation between the polarization vector and the magnetic field orientation. 
According the theory of polarized maser emission \citep{Goldreich1973},  
the direction of the linear polarization vector is  either parallel or perpendicular to the magnetic field (sky-projected) orientation, depending if the angle between the field orientation and the maser propagation direction (the so called Van Vleck angle, $\theta$) is less than or over 55$^{\circ }$, respectively. 

In order to assess weather  $\theta$ is less or more than 55\dg, we used a two-steps approach. 
First, we used the full radiative transfer method (FRTM) code developed by \citet{Vlemmings2006a} and based on the model by \citet{NedoluhaWatson92}, to determine $\theta$ (and associated errors) in each individual maser feature with linearly polarized emission (see Appendix~\ref{analysis}).
We then considered the entire range of values allowed by the estimated errors, from $\theta - \Delta \theta$ to $\theta + \Delta \theta$. In cases where $\theta - \Delta \theta >55$\dg,  the magnetic field is undoubtedly perpendicular to the linear polarization vector, whereas for $\theta + \Delta \theta <55$\dg\  it is undoubtedly parallel.  
In cases where $\theta - \Delta \theta <55$\dg\ and $\theta + \Delta \theta >55$\dg,  the magnetic field is either perpendicular or parallel to the linear polarization vector. In those cases, we assumed  that the magnetic field is more likely parallel if |$ (\theta + \Delta \theta) - 55^{\circ}| < | (\theta - \Delta \theta) - 55^{\circ}|$ (otherwise more likely perpendicular). 
It turned out that most of the  maser features with a relatively low (i.e. a few percent) fractional polarization fall in this last group, owing to larger negative error-bars with respect to positive error-bars (see Appendix~\ref{analysis} for an explanation).

As a second step, we plotted  the magnetic field vectors according to their $\theta$ angles calculated in the first step (these include all the linearly polarized features displayed in Fig.~\ref{pol_map}). 
We started analyzing the richest maser cluster, "a", which displays a remarkably consistent orientation of the linear polarization vectors (Fig.~\ref{pol_map}, inset {\it a}). We  noticed however that the presumed magnetic field orientations of adjacent masers changed by $\sim$90\dg\ in several cases. 
In fact, one limitation in our analysis is that the FRTM code overestimates $\theta$ for saturated masers \citep{Vlemmings2006a}, which implies that the condition of perpendicularity is erroneously satisfied in some cases.  Therefore, sudden changes of 90\dg\ in magnetic field orientation probed by adjacent masers may indicate that saturation effects are at play. 
Notably, when excluding maser features with \pl$\ge$5\%, we found that  the presumed magnetic field orientations of adjacent masers become  consistent with one another (within the errors). 
We interpret this result in terms of an {\em empirical threshold} of linear polarization fraction (\pl$=$5\%)   above which the water masers enter into the saturation regime, and therefore their polarized signal does not trace the magnetic field anymore.  
The goodness of this criterion is also demonstrated by the fact that 
it automatically excludes all the features in cluster "d" (except one), which indeed show a number of properties ascribable to saturation: 
 the highest intensity, including flaring features (up to 2000 Jy), among the highest values of brightness temperature, and an origin in a turbulent region interested by strong shocks. 
This may explain the exceptional high values of their linear polarization fraction ($>$10\%; see Table~\ref{features}) and the random distribution of   their linear polarization vectors (see inset {\it d} in Fig.~\ref{pol_map}), indicating that the  polarization does not reflect the magnetic field in the region.  

In summary, in order to avoid contamination from  saturated masers,  
in our analysis we adopted a conservative approach by excluding all maser features with \pl$\ge$5\%, which are either saturated or going towards a saturation state. 
A similar result was obtained in previous works by \citet{Vlemmings2006b} and \citet{Surcis2011a}, who  also suggested that a high linear polarization fraction, $P_l>5$\%, can only be produced when the maser is saturated.

Based on this analysis, we found that the magnetic field is parallel to the linear polarization vectors for most of the identified maser features with \pl$<$5\% (these have their $\theta$ angles in boldface 
in Table\ref{features}). 
Figure~\ref{B_map} shows the resulting (sky-projected)  magnetic field vectors  for the maser features with \pl$<$5\%,  overplotted on the  1.4~mm continuum map tracing the dust emission (same as in Figure~\ref{pol_map}),  as well as the 8.4~GHz continuum imaged with the VLA \citep{Wilner1999}. 
Interestingly, the magnetic field inferred from the \wat\,masers is on average oriented along E-W, i.e. along the mm dust and radio continuum emissions, suggesting a physical relation between them (see \S~\ref{B_dyn}).  
The fact that the field orientations for neighboring maser features are mostly consistent with one another, increases our confidence in the physical relevance of the inferred magnetic field structure. 

%-------------------------------------------------------------
\subsection{Circular polarization: magnetic field strength}
\label{circ}
%-------------------------------------------------------------
Besides linear polarization, we also detected circularly polarized emission toward ten maser features, varying in the range $P_{\rm{V}}=0.2-1.6\%$. 
Considering only  masers with \pl$<$5\%, 
i.e. 037, 051, 063, 104, and 105 (all towards component "A" of the mm continuum), we obtain values of the magnetic field strength along the l.o.s.,  $B_{\rm{l.o.s.}}$, of (--60$\pm$11) mG, (--84$\pm$35) mG, (+360$\pm$120) mG, (+169$\pm$34) mG, and (+161$\pm$41) mG, respectively\footnote{A negative magnetic field strength indicates that the magnetic field is pointing toward the observer; positive away from the observer.}. 

\citet{Sarma2002} also conducted  Zeeman measurements in the 22 GHz \wat\,masers with the VLA (beamsize $\sim$0\pas1),  yielding a field strength $B_{\rm{l.o.s.}}= 42\pm3$~mG, much lower than the values we quote here.  
However, their measurements used only the strongest feature at --49.1~\kms, located  at $\alpha(J2000) = 02^h 27^m 04\rlap{.}^s866$,   $\delta(J2000) = +61^{\circ} 52' 24.89"$, i.e. in the cluster "d" of \wat\,masers (Fig.~\ref{pol_map}, inset "d"), while our measurements of circular polarization include only cluster "a". 
We also measured a Zeeman splitting for the same feature detected with the VLA (corresponding to maser ID 018 in Table~\ref{features}), yielding $B_{\rm{l.o.s.}}=59\pm9$~mG: this is consistent with the value inferred by \citet{Sarma2002}\footnote{The small discrepancy between the VLBI and VLA values may be due to blending with other features within the VLA beam (e.g., features 018 and 019), as also noticed by \citet{Sarma2002} based on the asymmetry in the maser line profile, which required multiple components for a satisfactorily  Gaussian fit.}.  
We did not however include this feature (nor any other strong maser feature in cluster "d") in our analysis because it does not satisfy the condition that the masers should be unsaturated, and therefore  cannot provide reliable estimates of the magnetic field strength. 
On the other hand, we fear that a measurement of the Zeeman splitting towards our cluster "a" with the VLA may have been prevented by
spatial blending of features with  positive and negative components  of $B_{\rm{l.o.s.}}$ with similar strengths within the VLA beam (one example is provided by  features 036 and 039).

The total amplitude of the magnetic field can be estimated with $B = B_{\rm{l.o.s.}}/cos(\theta)$, yielding  $B_{\rm{037}}=-95^{-20}_{+45}$~mG, $B_{\rm{051}}=-128^{-57}_{+77}$~mG, $B_{\rm{104}}=+295^{+214}_{-30}$~mG, and $B_{\rm{105}}=+322^{+256}_{-37}$~mG\footnote{The errors in {\it B} take into consideration the errors of both $B_{\rm{l.o.s.}}$ and $\theta$.} (we excluded the maser component 063 because  $\theta$ is $\sim$90\dg, and $B$ cannot be constrained). 
The locations of the maser features for which we measured the Zeeman splitting, along with the associated values of the magnetic field strengths, are plotted in Figure~\ref{B_map} and reported in Table~\ref{B:table}. 

It is perhaps worth noting that the amplitude of the magnetic field strength decreases with distance from the radio and mm continuum peaks in component "A", and has opposite signs on either side of the continuum peak (see Fig.~\ref{B_map} and Table~\ref{B:table}). In \S~\ref{B_dyn} we show that the continuum peak locates the protostellar position and the radio continuum emission comes from the protostellar jet.  
 Therefore, this finding may indicate that the magnetic field decreases as a function of distance from the protostar along the jet axis, probing both the redshifted and blueshifted lobes of the molecular outflow.

%-----------------------------------------------------------------------------
\begin{table}
  \renewcommand\thetable{2}
\caption{Measurements of the magnetic field strength from \wat\,masers with Zeeman splitting in TW-A. 
}              
\label{B:table}       
\centering  
\small
\begin{tabular}{cllcc} 
\hline\hline                  
\noalign{\smallskip}
\multicolumn{1}{c}{Maser}  &RA (J2000) & DEC (J2000) &  \multicolumn{1}{c}{B strength} & \multicolumn{1}{c}{R$^{a}$} \\
 \multicolumn{1}{c}{ID}   & \multicolumn{1}{c}{(h:m:s)} &  \multicolumn{1}{c}{($^\mathrm{o}$:':")} & (mG) & (arcsecond)\\
\noalign{\smallskip}
\hline
\noalign{\smallskip}  
037   & 02:27:04.7444 & 61:52:24.694 & $-95^{-20}_{+45}$ & 0.347 \\
051   & 02:27:04.7409 & 61:52:24.705 &$-128^{-57}_{+77}$ & 0.314 \\
104   & 02:27:04.6994 & 61:52:24.655 &$+295^{+214}_{-30}$ & 0.110 \\
105   & 02:27:04.6993 & 61:52:24.656 &$+322^{+256}_{-37}$ & 0.111 \\ 
\noalign{\smallskip}
\hline    
\end{tabular}
\tablefoot{(a) The (sky-projected) separation between the masers with Zeeman splitting measurements and the putative location of the TW-A protostar, estimated from the peak of the 8.4~GHz continuum, $\alpha_{\rm{2000}}=02^{h}27^{m}04^{s}\!.7103$ and $\delta_{\rm{2000}}=+061^{\circ}52'24''\!\!.650$.}
\end{table}
%--------------------------------------------------------------------------------------
\begin{figure*}
\centering
\includegraphics[width=\textwidth]{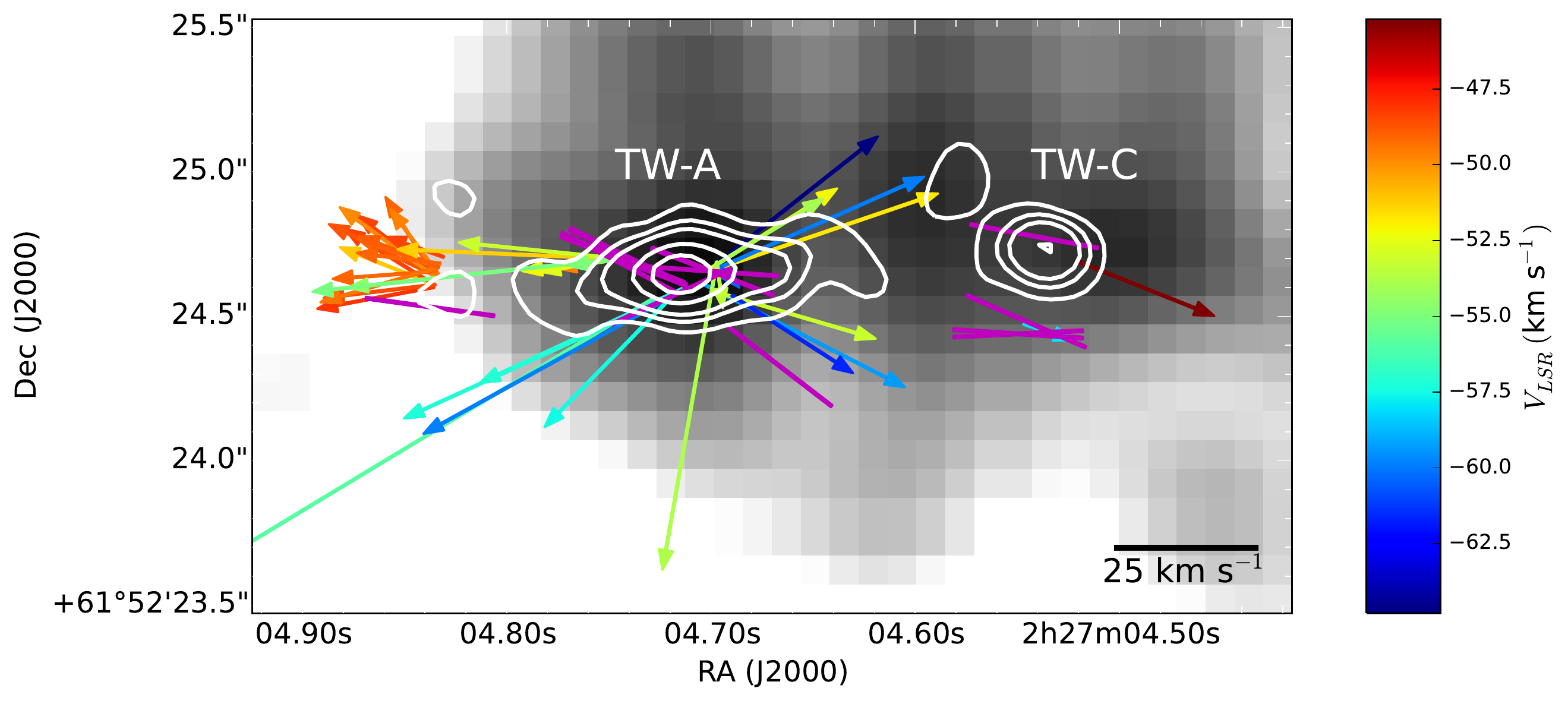}
\includegraphics[width=\textwidth]{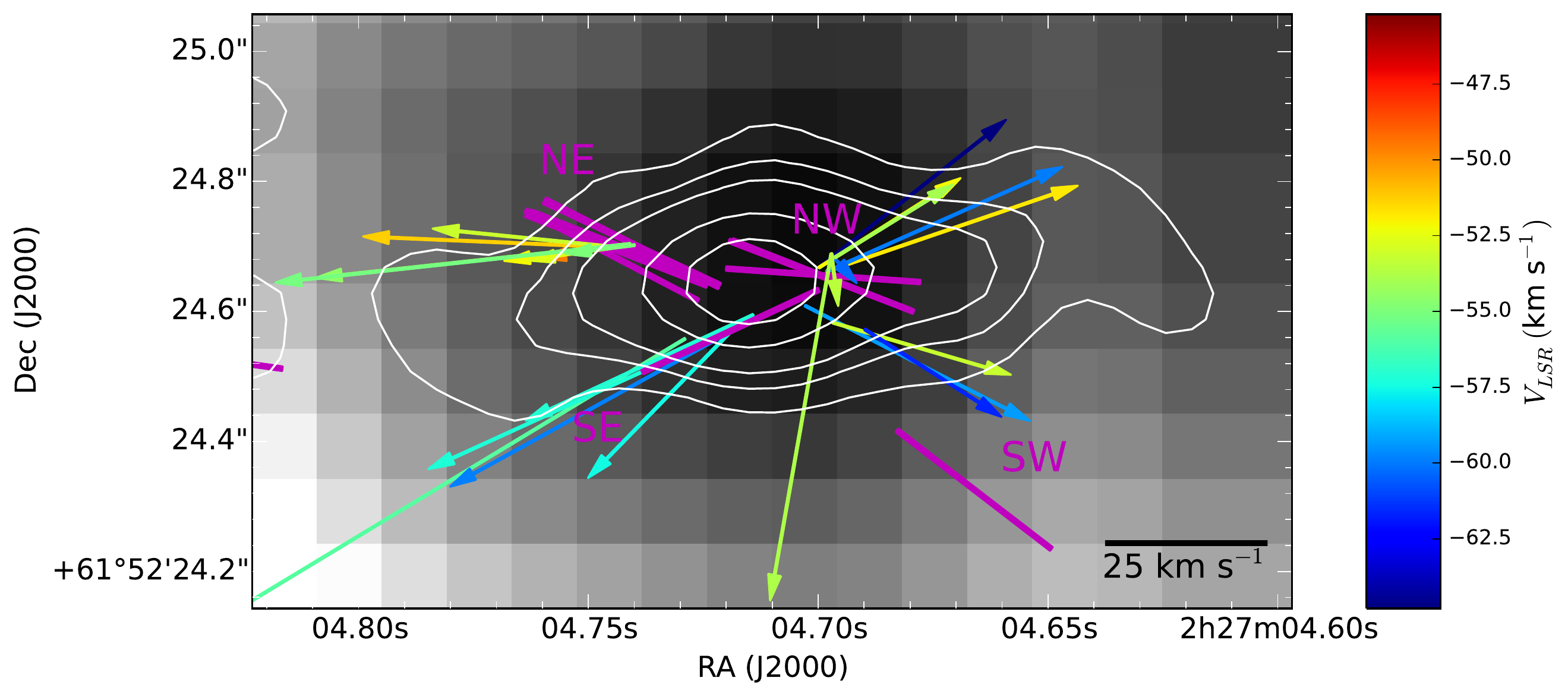}
\caption{Comparison between the magnetic field orientations of \wat\,masers (purple segments) measured in this work    and proper motions of \wat\,masers (arrows) measured by \citet{Hachi06}. 
The 8.4 GHz emission imaged with the VLA by \citet{Wilner1999}  (white contours) is overploted onto the 1.4~mm continuum emission mapped with the PdBI by \citet{Wyrowski1999} 
 (gray scale). 
Colors denote  l.o.s.  velocity (color scales on the right-hand side) and   
 the scale for the proper motion amplitude is given in the lower right corner (both in  \kms). 
 ({\it Top panel})  
The outward proper motions of the masers to the east and west  suggest the presence of a collimated bipolar molecular outflow along E-W (the axis of the synchrotron jet), whereas the proper motions of the masers closer to the radio continuum peak identify a wider-angle biconical outflow  driven by the the TW-A high-mass protostar.
The western component identifies TW-C, the binary companion of TW-A \citep[e.g.,][]{Wyrowski1999,Zapata2011}. 
 ({\it Bottom panel})   
Expanded view towards the core of the synchrotron jet, showing more in details
the relation between the \wat\,masers magnetic field orientations  (purple segments)  and proper motions (arrows). There are four main knots of masers, located towards NE, NW, SE, and SW,  with respect to the radio continuum peak (labeled  accordingly in the figure). 
The  proper motions identify a biconical, bipolar molecular outflow driven by the the TW-A high-mass protostar. Note  the misalignment between magnetic field and velocity vectors, particularly in the NE and NW knots (see \S~\ref{shocks} for an explanation).  
}
\label{fig:PM}
\end{figure*}
%--------------------------------------------------------------------------------------

%==============================================================================
\section{Discussion}
\label{disc}
%==============================================================================
One relevant  question in HMSF is the relation between the gas dynamics and the magnetic field in regulating mass-accretion and mass-loss.  
VLBI measurements of \wat\,masers can provide a detailed description of gas kinematics and magnetic field structure  on scales from tens to hundreds AU,  which are the smallest accessible scales in studies of HMSF, and have therefore the potential to address such an open question.   
 \w3\,contains the best known (archetypical) case of  synchrotron jet driven by an embedded high-mass YSO and associated with \wat\,masers,  and therefore provides a good target for  investigating this issue. 
\citet{Alcolea1993}, and later on \citet{Hachi06}, performed multi-epoch VLBI observations of   H$_2$O masers, thus probing the 3D velocity field of the molecular gas surrounding the synchrotron jet on linear scales of hundreds of AU. 
Our polarimetric measurements of \wat\,masers enabled us to infer the magnetic field  strength and orientation  on similar linear scales, thus providing the missing piece of information to help us assess the above-mentioned relation. 

%----------------------------------------------------------------------------
\subsection{Gas dynamics and  magnetic field structure}
\label{B_dyn}
%---------------------------------------------------------------------------- 

Figure~\ref{fig:PM} shows an overlay of the dust emission from the hot-core \citep[greyscale;][]{Wyrowski1999}, synchrotron emission from the radio jet \citep[white contours;][]{Wilner1999},  water maser proper motions (arrows) measured by \citet{Hachi06}, and the direction of magnetic field vectors as determined from our polarization measurements (purple segments). 
This overlay illustrates a number of physical properties of the system that we describe in detail below. 

The first most notable element is the presence of multiple components in the radio continuum, as already noticed by \citep{Wilner1999}: 
a central strongest component, and two secondary peaks to the west (stronger)  and to the east (fainter), respectively.
 The strongest component is centered at the mm dust continuum peak "A" (see also Fig.~\ref{pol_map}) and 
has an elongated morphology along E-W: this is  the synchrotron jet discovered by \citet{Reid1995}. 
Considered their alignment  with the  axis of the central jet, the secondary components  were initially both interpreted as Herbig-Haro objects  resulting from strong shocks produced as the fast protostellar jet impinges onto the dense ambient medium. 
While this may explain the (weaker) eastern component, an alternative interpretation has been suggested for the western (stronger) component, which is nearly coincident with  the dust continuum peak "C",  and shows a rising spectral index ($\alpha=+0.9$ between 8 and 15~GHz - \citealt{Wilner1999}),  as expected from an  ionized stellar wind and/or thermal radio jet. 
 \citet{Wyrowski1999} used several $K$ components of the HNCO (10--9) transition to map the rotational temperature of the molecular gas in \w3, and found evidence for two heating sources in the region (with $T_k \sim$200~K), coincident with the dusty peaks A and C.  
  These findings are  interpreted as evidence for a binary made of two high-mass (B-type) protostars embedded in the \w3\ hot molecular core  \citep{Zapata2011}. In the remaining of this paper, we indicate these protostars as TW-A and TW-C (these are labeled in  the top panel of Figure~\ref{fig:PM}). 

The second element worth discussing is the kinematics traced by \wat\,masers. 
They are mainly concentrated towards the three components of the radio continuum, 
along E-W,  with  the eastern and western maser clusters showing outward  proper motions at about 20~\kms.  
The natural interpretation is that 
the \wat\,masers probe the molecular component of the synchrotron jet \citep[e.g.,][]{Hachi06}. 
In fact,  a closer look at the measured proper motions in different clusters reveals  additional features.
First,  
there is a strong asymmetry  between  the eastern and western maser clusters.  
 The eastern cluster contains the strongest, saturated masers (with peak flux densities up to about 2000~Jy), 
 and shows a remarkably linear structure perpendicular to its motion (see also Fig.~\ref{pol_map}, inset "d"), reminiscent of a bow-shock or a propagating exciting front (similar to what seen in  other \wat\,maser sources: e.g., G24.78; \citealt{Moscadelli2007} and Cepheus~A; \citealt{Torrelles2001}): this structure can be a component of the same  jet propagating from TW-A. 
 On the other hand,  the western component excites a much smaller number of (weaker) masers, which seem to move consistently towards south-west: these masers are most probably  associated with the second YSO, TW-B. 
Finally, in the central cluster (labelled "a" in Fig.~\ref{pol_map}), the masers  show a {\it biconical} expanding flow centered on the presumed location of the exciting high-mass protostar, TW-A, with an opening angle of about 15--30\dg\ (estimated from the proper motion position angles) and with somewhat higher velocities (up to 90~\kms) with respect to the external components. 
This suggests that the outflow from the TW-A object has potentially two different components, a collimated synchrotron jet elongated E-W across a few thousand of AU and a somewhat wider-angle biconical molecular flow at its center  (a similar outflow structure is observed in the high-mass Cepheus A HW2 system; \citealt{Torrelles2011}).

The third  feature displayed in Figure~\ref{fig:PM}  (top panel), is the apparent consistency between gas kinematics and magnetic field orientation, 
at least on scales of order of one thousand AU or larger, comparable with the size of the dusty core. 
Indeed, the average magnetic field orientation  inferred from the \wat\,masers is   mainly  E-W, i.e. well aligned with the axis of the  synchrotron jet as well as with the axis the molecular outflow (the latter can be defined by the average orientation of the maser proper motions). 
This is also apparently consistent with the  dust polarized emission, mapped with the SMA at 0.9~mm (1\pas5 beamsize; \citealt{Chen2012}), which shows a magnetic field  aligned with  the E-W elongation of the dusty core surrounding the TW-object(s).  
This finding is however surprising since the mm emission traces {\it quiescent} material in the dusty hot core, while \wat\, masers trace {\it shocked} material in the prostostellar jet.
Therefore, the two probes are not expected to provide  necessarily consistent pictures for the magnetic field. 

The much higher angular resolution provided by the VLBI measurements, with respect to imaging of   polarized emission from dust, enables us to take a much closer look at the relation between gas kinematics and magnetic field morphology.  
 The bottom panel of Figure~\ref{fig:PM}  offers  a zoomed view of the innermost parts of the synchrotron jet within 1500 AU from TW-A.  
The first notable feature is the presence of four different knots of masers, located towards NE, NW, SE, and SW with respect to the radio continuum peak, respectively (these knots are labeled  accordingly in the figure). 
Their proper motions identify a clear biconical expanding flow (as already noticed). 
Since we also determined the magnetic field orientation for some of the polarized masers associated with  these four different knots, we can directly compare it with the velocity vectors. 
One caveat is that, since the multi-epoch proper motion and the polarimetric experiments were conducted 5--6 years apart, a {\it one-to-one} comparison between magnetic field and proper motion orientations in individual  maser features is not possible. 
Nevertheless, we can still compare their {\it average} relative orientations  at different locations. 
We find that the angles representing the minimum differences  between the average position angles of proper motions and magnetic fields in the plane of the sky are respectively: 
6\dg\ for SE, 
14\dg\ for SW,  
30\dg\ for NE, and
54\dg\ in the NW\footnote{These are the values for the angle $\phi$ defined in \S~\ref{B_V} and Appendix~\ref{B_geometry}.}.  
This finding indicates that, on scales of few hundreds AU, there is a significant misalignment between the orientations of the magnetic field  and the velocity vectors. 
The origin of such a misalignment can be  understood with an origin of water masers  in fast shocks, as we detail in next section.

%------------------------------------------
\subsection{Magnetically supported shocks}
\label{shocks}
%------------------------------------------

Water masers are thought to emerge behind fast C- or J-type shocks \citep{Hollenbach1979,Elitzur1989,KaufmanNeufeld1996,HEM13}. 
We show here  that   the magnetic field properties (strength and  orientation)  inferred from the  observed polarized water maser emission  
are indeed consistent with an origin in (magnetically supported) shocks.

 As a shock propagates in the ambient medium, it alters the initial magnetic field configuration in the circumstellar gas; in particular,  the magnetic field component perpendicular to the shock velocity (and parallel to the shock front) is compressed and, therefore, enhanced with respect to the parallel component  (which remains unaffected).  As a consequence, we expect the magnetic fields probed by the \wat\,masers  to be   along the shock front\footnote{Observational biases also favor the same component because of maser path length, which will favor the compressed field perpendicular to the velocity and parallel to the shock.} (see quantitative discussion in \S~\ref{B_V}).
 This implies that the \wat\,masers  provide the  structure of the magnetic field in the post-shock gas, rather than in the "quiescent" circumstellar gas, and therefore they are a good probe of the shock morphology.  
 Nevertheless, when the information on the measured orientation and strength of the magnetic field (in the post-shock gas) is coupled with 
 the knowledge  of the maser velocities, 
 we can still obtain constraints on the strength and geometry of the  magnetic field in the pre-shock circumstellar gas (as  we detail  in \S~\ref{B_preshock}).

%--------------------------------------------------------------------
\subsubsection{Geometry of the magnetic field in the post-shock gas}
\label{B_V}
%--------------------------------------------------------------------
After the shock passage, the component of the magnetic field parallel to the shock surface, and therefore perpendicular to the shock velocity, is expected to become dominant  in the post-shock gas. 
This implies a well defined prediction on the expected relative orientation between the gas motion and the magnetic field.   
 In the assumption that the  proper motions of water masers closely represent the shock velocity\footnote{Since the proper motions dominate over  the line-of-sight velocity, here we assume that the shocks responsible for the excitation of \wat\,masers move in the plane of the sky.}, we would  expect an angle of 90\dg\ between  the magnetic field and the proper motion orientations. 
However, as pointed out in \S~\ref{B_dyn},  the observed angles between the sky-projected component of  the magnetic field and the proper motions are  significantly less than 90\dg\ (see also  the bottom panel of Figure~\ref{fig:PM}). 
A natural explanation for the apparent deviation from the expected perpendicularity between magnetic field and velocity vectors is conductible to  3-D effects.  

In Appendix~\ref{B_geometry}, we show that the ratio between the components of the magnetic field parallel, $B_{\rm{||}}$, and perpendicular, $B_{\perp}$, to the proper motion orientation can be expressed as follows:  
\begin{equation}
\label{Bratio}
\centering
$$ B_{\perp} / B_{\rm{||}} = \sqrt{\cot^2(\theta) / cos^2(\phi) + \tan^2(\phi)} $$
\end{equation}
where  $\theta$ is the angle between the magnetic field vector {\bf B} and the line of sight (as already defined in \S~\ref{lin_pol}), and $\phi$ is the angle between the sky-projected magnetic field  and the proper motion orientation (whose measured values are given at the end of \S~\ref{B_dyn}). 

This ratio can provide a measure of the relative orientation of the magnetic field with respect to the proper motions.  In a first approximation,  if $B_{\perp} > \ {\rm or} \gg B_{\rm{||}}$, then the magnetic field  can be considered most probably perpendicular to the proper motions. 
In order to have \  $ B_{\perp} > B_{\rm{||}} $, $\phi$ should be large and $\theta$ should be small. 
Although we find $\phi \le$30\dg\ in three (out of four) of the {\it knots} with measured proper motion and magnetic field orientations (the NW knot  has $\phi$=54\dg; see end of \S~\ref{B_dyn}), those masers present a large negative errorbars in the measured values of \ $\theta$ (see the boldface numbers reported in col.~13 of Table~\ref{features}), allowing values as low as \ 15\dg$\le \theta \le$30\dg. As an example, using Equation~\ref{Bratio}  and taking the maximum value in the range of \ $\theta = $30\dg \ and a minimum value of \ $\phi = $10\dg, we obtain a minimum value of \ $ B_{\perp} \approx  1.8 \, B_{\rm{||}}$, implying a lower limit of \ $\approx$60\dg\ for the angle between the magnetic field orientation and the maser velocity (i.e., the shock velocity).
On the other hand, taking a minimum allowed value of $\theta \approx$15\dg\ and a maximum value of \ $\phi \approx$50\dg, it would imply a maximum ratio \ $ B_{\perp} / B_{\rm{||}} \approx$6,  corresponding to an angle of \ $\approx$80\dg\ between the magnetic field and the shock velocity.
 
This simple analysis shows that, the observed misalignment between the magnetic field and the velocity vectors in the plane of the sky, can be explained with an origin in magnetically supported shocks, where (the post-shock) magnetic field is expected to be  perpendicular to the maser (and shock) velocity. 

{A caveat is that our analysis is based on  measurements of magnetic fields and velocities that were  {\it not simultaneous}, which prevented us to compare the two properties within individual maser features.
Therefore the observed relation between  magnetic field and velocity vectors is confirmed on scales of the identified maser knots, that is several tens of AU.
 Future, simultaneous observations of water maser polarization and proper motions will allow us to verify if such a relation holds from the smallest scales of individual maser features to scales of the entire  outflow.
 The power of simultaneous observations of polarized emission and proper motions  has been recently demonstrated in the case of 6.7~GHz methanol masers by \citet{Sanna15}, who   assessed the existence of  a coupling between the magnetic field and the motion of the circumstellar gas. 
 Unlike methanol, water masers are  shock-excited, therefore establishing the same relation is  less straightforward in the case of water masers. 
We discuss this aspect in  \S~\ref{B_preshock}.

%---------------------------------------
\subsubsection{Constraints on the strength and geometry of the magnetic field in the pre-shock gas}
\label{B_preshock}
%---------------------------------------

While  \wat\,masers provide direct information on the (magnetic and kinematic) structure of the shock,  polarization and proper motion measurements (possibly simultaneous)  can be used together to get constraints on the magnetic field properties in the pre-shock gas as well. 

In order to relate the magnetic field strength in the pre- and post-shock region, we can  adopt the J-shock model of \citet{Hollenbach1979}. 
In particular, for a magnetically supported shock, \citet{Hollenbach1979} showed that by equating  the ram pressure of the pre-shock gas and the  magnetic pressure of the post-shock gas, one obtains (their eq. 2.34): 
\begin{equation}
$$\left(\frac{B_{0\perp}}{10^{-3} \ \rm mG}\right) \sim 77 \  \left(\frac{v_s}{100~\rm km s^{-1}}\right) \ \frac{n_0^{3/2}}{n_s}$$ 
\end{equation}
where $B_{0\perp}$ is the component of the field compressed by the shock (perpendicular to its propagating direction), $v_s$ is the shock velocity, $n_0$ and $n_s$  are the pre-shock and post-shock gas densities\footnote{Note that the numerical factor is not scalar, but absorbs the unit of the ratio of densities with different exponents.}, respectively. 
By assuming a pre-shock density of $10^7$~\cmc\ \citep{Wyrowski1999,Zapata2011} and typical shock velocities of 50--100~\kms, a post-shock density of $10^9$ \cmc\ (required for water maser action - \citealt{HEM13}) can be obtained for a pre-shock magnetic field of a $\sim$1--2 mG\footnote{A model for C-type shocks provides magnetic fields of the same order of magnitude (see Eq.~4.5 of \citealt{KaufmanNeufeld1996}).}.  
According to the J-shock model of \citet{Hollenbach1979}, 
 the  component perpendicular to the shock propagation (parallel to the shock front) is enhanced by a factor equal to the ratio between the post- and pre-shock densities (their Eq. 2.21): $B_{\perp}  = B_{0\perp} \ n_s/n_0$.
Therefore, after the shock passage,  $B_{\perp}$ would be amplified to  100--200 mG (assuming $n_0 = 10^7$~\cmc, $n_s = 10^9$~\cmc, and $B_{0\perp}=1-2$~mG). 
This is consistent with the range of values  of the  magnetic field strength measured with the \wat\,masers (Table~\ref{B:table}).  

Since according to the same model the component of the magnetic field parallel to the shock velocity does not vary (see eq. 2.16 in \citealt{Hollenbach1979}),  
we can use this prediction to derive constraints on the strength of the parallel component in the preshock gas.  The analysis in \S~\ref{B_V} indicates a maximum ratio \ $ B_{\perp} / B_{\rm{||}}  \approx$6 in the post-shock region. Since $B_{\perp}$ \ is likely amplified by a factor of 100 in the fast shocks producing the water masers while \ $B_{\rm{||}}$ \ remains unchanged, we can infer that 
\ $ B_{0\rm{||}} \ge 10 \, B_{0\perp}$. 

According to this crude analysis, we constrain the magnetic field strength in the pre-shock gas  to 10--20~mG. 
This range is the right order of magnitude for the magnetic field derived from the synchrotron emission model by \citet{Reid1995}, which infers a strength of the magnetic field  $B_{\rm{0}}=10$~mG\footnote{Note that this value  is constrained  to no better than a factor of a few from the fit of the radio continuum data.}.   
Likewise, our estimate of magnetic field strength is also consistent with the value of $B_{\rm sky}=17$~mG reported by \citet{Chen2012} from dust emission. 

We stress that, even allowing a factor of 10 between    $B_{0\rm{||}}$ and $ B_{0\perp}$, $B_{\perp}$ would still dominate in the post-shock gas, consistent with an estimated range of 100--300~mG.

In summary, the magnetic field would evolve from having a dominant  component parallel to the shock velocity in the pre-shock gas, with field strengths of the order of a few tens of mG,  to being mainly dominated by  the perpendicular component of order of a few hundred of mG in the post-shock gas exciting the \wat\,masers\footnote{The parallel component could dominate in the post-shock gas only if it   were $>$100~mG in the pre-shock gas, i.e. 2 orders of magnitude stronger than the pre-shock perpendicular component; a 100 mG field in $10^7$~\cmc\ gas is however implausibly high. Therefore, we can rule out that with \wat\,masers we are measuring the parallel component  and we are dominated by the perpendicular component.}. 
The general implication of this finding is that in the  undisturbed (i.e. not-shocked) circumstellar gas, the flow velocities would follow closely the magnetic field lines, while in the shocked gas the magnetic structure would be re-configured to be parallel to the shock front.

%===================================
\section{Summary and Conclusions}
%===================================

We present the first VLBI polarization measurements of \wat\,masers in \w3.
This HMSF region contains the best known (archetypal) case of a synchrotron jet driven by an embedded high-mass YSO, the TW-object or TW-A, 
 and therefore provides a good target for  investigating  the relation between the gas dynamics and the magnetic field in regulating mass-loss in HMSF.  
In particular, our polarimetric measurements of \wat\,masers  have enabled us to infer the magnetic field  strength and orientation, that can  be directly  compared to the kinematics of the molecular outflowing gas, on scales of order of tens to hundreds AU,  among the smallest accessible scales in studies of HMSF.  

\vspace{0.1 cm} 

The main findings of this work are the following:
\begin{enumerate}[1.]

\item
We measured a linear polarization fraction varying in the range 0.9--42\%, making \w3\ the highest polarized \wat\,maser source on  VLBI scales known in the Galaxy. 
The masers show also circularly polarized emission, in the range 0.2--1.6\%. 

\vspace{0.2 cm} 
\item 
On scales of order of one thousand AU or larger, 
the average magnetic field orientation probed by \wat\,masers is mainly  E-W,  well aligned with the axis of the synchrotron jet driven from the TW-A protostar, suggesting that the molecular masers may probe the  magnetic field in the protostellar jet. 

\vspace{0.2 cm} 
\item
On smaller scales, 10s to 100s of AU, a detailed comparison between the magnetic field orientations and the proper motions of the \wat\,masers observed at the center of the protostellar jet/outflow, 
reveals a misalignment between the magnetic field and the velocity vectors, which can be explained with an origin in magnetically supported shocks. 

\vspace{0.2 cm} 
\item
Since water masers emerge behind  C- and/or J-type shocks, the shock passage alters the initial magnetic field configuration in the circumstellar gas, 
by compressing and enhancing the component of the magnetic field perpendicular to the shock velocity with respect to the parallel component  (by a factor equal to the ratio between the post- and pre-shock densities:  typically a 100). 
In the  gas shocked by the synchrotron jet, we estimate a total field strength  in the range $\sim$100-300~mG  (at densities of $10^9$~\cmc) 
and we conclude that fields of this order of magnitude are expected if the  observed polarized water masers indeed emerge behind magnetically-supported shocks which compress the field component along the shock front. 

\vspace{0.2 cm} 
\item
Although the water maser polarization measurements  alone cannot  provide
a {\it direct} measurement of the magnetic field properties in the {\it quiescent} (pre-shock) circumstellar gas, nevertheless by combining the information on the  orientation and strength of the magnetic field (in the post-shock gas)  with  the knowledge  of the maser velocities, 
we could constrain  the magnetic field strength in the pre-shock circumstellar gas to 10--20~mG  (at densities of $10^7$~\cmc), which is consistent with previous estimates from a synchrotron jet model and dust polarization measurements. 

\vspace{0.2 cm} 
\item
We estimate a lower limit of 10  for the ratio of  the parallel and perpendicular components of the magnetic field  with respect to the proper motions in the pre-shock gas. This indicates that  the flow velocities follow closely the magnetic field lines and is suggestive of  a local coupling between the magnetic field and the kinematics of the circumstellar gas in the \w3\,core and the protostellar jet from the TW-object.

\end{enumerate}

This study demonstrates that, although \wat\,masers can naturally probe only the magnetic field in  regions shocked  by protostellar jets, by combining the knowledge of the 3D velocities and magnetic field orientation/strength, it is still possible to derive constraints on the properties of the magnetic field in the pre-shock circumstellar gas at the base of  protostellar outflows. 
In particular, our results in \w3\ suggest the presence of a local coupling between the magnetic field and the gas kinematics, indicating that magnetic fields can be dynamical important  in driving the gas outflowing from a high-mass protostar. 
Future studies based on VLBI multi-epoch  observations of \wat\,masers in full polarisation mode 
(which  naturally provide magnetic field and velocity vectors in the same maser features), 
 have the potential to open up a new window into the investigation of the role of magnetic fields in the gas dynamics of high-mass protostellar disk-jet systems, 
on scales  of only tens to hundreds AU from the exciting massive protostar. 

\vspace{0.3cm}
%----------------------------------------------------------------------------------------
\begin{acknowledgements}
We are grateful to Dr F. Wyrowski and Dr  K. Hachisuka for providing the 1.4~mm and the 8.4 GHz continuum images, respectively. We are grateful to Dr K. Hachisuka for useful discussion on absolute proper motion measurements of water masers. W.V. acknowledges support from ERC consolidator grant 614264. 
\end{acknowledgements}

\input{table_paper_inp}

\bibliographystyle{aa}
\bibliography{biblio}  

\appendix
\section{Analysis of Maser Polarization Data}
\label{analysis}

 We analyzed the polarimetric data following three main steps:
 i)  identification of individual \wat~maser features and measurement of 
their linear polarization fraction and polarization angle; 
 ii) determination of the orientation of the magnetic field on the plane of the sky; 
 iii) measurement of Zeeman splitting and magnetic field strength along the l.o.s.. 
 
\vspace{0.1 cm} 
The methodology is described in full detail in \citep{Surcis2011a, Surcis2011b}.
Here below we describe the main steps for completeness. 

\vspace{0.2 cm} 

 i) 
To identify the water masers in the W3(H$_2$O) region, 
first we searched for maser emission channel by channel, with a signal-to-noise ratio greater than 10-sigma noise level;  
then, we fitted the identified maser spots with a 2D Gaussian function (using the AIPS task IMFIT); 
and finally we defined individual maser features as collections of three or more maser spots  in consecutive velocity channels within the full width half-maximum (FWHM) beamsize. 
Once identified all the maser features, the next step was to  estimate the linear polarization fraction and the polarization angle in each of them. 
For this purpose, we produced spectra of total intensity (\textit{I}), polarized intensity (\textit{POLI}), and polarization angle (\textit{POLA}) 
(see \S~\ref{obssect}) 
at the position of each identified maser feature (integrating in a box $3\times 3$ pixels). 
The linear polarization fraction ($P_{\rm{l}}$) and the linear
polarization angle ($\chi$) of a single  feature are determined taking the average values in 
 consecutive channels across its total intensity spectral profile with $POLI\ge 5\sigma$ (Figure~\ref{POL_plots} shows a few examples).
Table 1 reports the main parameters of all the maser features identified in the region.

\vspace{0.2 cm} 

ii) The orientation or position angle of the magnetic field on the plane of the sky can be determined using the theory of polarized maser emission, 
as outlined in \citet{Goldreich1973}. 
In particular, 
the relation  between the measured polarization angle $\chi$ and the magnetic field angle on the sky  
depends on the angle between the maser propagation direction and the magnetic field orientation, $\theta$, 
with the linear polarization vector perpendicular to the field for $\theta>\theta_{\rm{crit}}=55$\d, 
where $\theta_{\rm{crit}}$ is the so-called Van Vleck angle, and parallel otherwise \citep{Goldreich1973}. 
 $\theta$ can be determined from the emerging brightness temperature, 
 \tbo\footnote{The emerging brightness temperature, \tbo, is defined as the product of the brightness temperature, $T_{\rm{b}}$,  
 and the solid angle of the maser beam, $\Delta \Omega$.}, and 
 the fractional linear polarization, $P_{\rm{l}}$, of the \wat\, maser emission   \citep{Goldreich1973}. 
\tbo\,is estimated by fitting the spectral profiles of individual water maser emissions by using the FRTM code developed by \citet{Vlemmings2006a} and based on the model of  \citet{NedoluhaWatson92}.   
In this model, the spectral profiles of the total intensity,
 linear polarization, and circular polarization  depend on $T_{\rm{b}}\Delta\Omega$, 
 as well as  the intrinsic thermal linewidth, \dvi \footnote{\dvi\, is basicaly the FWHM of the Maxwellian distribution of particle velocities.}. 
 The FRTM code uses a $\chi^{2}$-fitting analysis and provides in output \tbo\, and \dvi\,(refer to \citealt{Surcis2011a} 
 for a detailed description).  
From  $P_{\rm{l}}$ and \tbo\,   
we can then determine $\theta$ (the estimated values are reported in col. 13 in Table~\ref{features}).
We note that the estimated $\theta$ values have large negative (and small positive) error bars. This is a consequence of the probability calculation in the FRTM code. In particular, \citet{Vlemmings2006a} calculated the expected relation between $\theta$ and $P_{\rm{l}}$ for different values of \tbo\, and showed that  there are typically three values of $\theta$  that produce the same linear polarization. For instance,  looking at their figure~6, one could realize that a maser with $\theta \sim$50\dg\ (for a non-saturated maser) would have the same $P_{\rm{l}}$ as a maser with $\theta \sim$10\dg\ or a maser with $\theta \sim$58\dg\ (in which case the linear polarization angle on the sky flips). 
Therefore, for a given (low) saturation level and low polarization level, it is more likely to have a component with $\theta <$55\dg, but the uncertainty interval toward the low end is very large.

\vspace{0.2 cm} 

iii)  The magnetic field strength along the l.o.s. can be calculated from the Zeeman-splitting measurements by using
$B_{\rm{l.o.s.}}={\Delta V_Z}{\alpha_Z}$, 
where $\alpha_Z$ is  the  Zeeman-splitting  coefficient. 
This is done by fitting the total intensity and circular polarized spectra 
of the \wat~masers identified in the high-spectral resolution {\it I} and {\it V} cubes (see \S~\ref{obssect} and Figure~\ref{Vfit} for a few examples). 
The best estimates of \tbo ~and \dvi\,(as described in step ii) are included in the FRTM code to produce the $I$ and $V$ models used 
in the fitting procedure. 
The Zeeman splitting depends on $T_{\rm{b}}\Delta\Omega$, the FWHM of the maser total intensity profile, \dvl, 
and the circular polarization fraction, $P_{\rm{V}}$. 
Once the field strength along the l.o.s. is estimated, one can determine the total magnetic field  strength   from $B=B_{\rm{l.o.s.}}/\cos{\theta}$. 

\vspace{0.2 cm} 

A final element worth noting is that the degree of saturation affects the fractional linear polarization $P_{\rm{l}}$. 
When the saturation sets in, the maser lines start to broaden, and the observed linewidth \dvl\,can become as large as \dvi.
For a saturated maser, \tbo\,and \dvi\,cannot be  properly disentangled by the model of \citet{NedoluhaWatson92}, and the FRTM code only provides a lower limit for \tbo\, and an upper limit for 
\dvi. 
As a result, the fitted values of $\theta$ are overestimated. 

%--------------------------------------------------------------------------------------
\begin{figure*} 
\centering
\includegraphics[width = 6 cm]{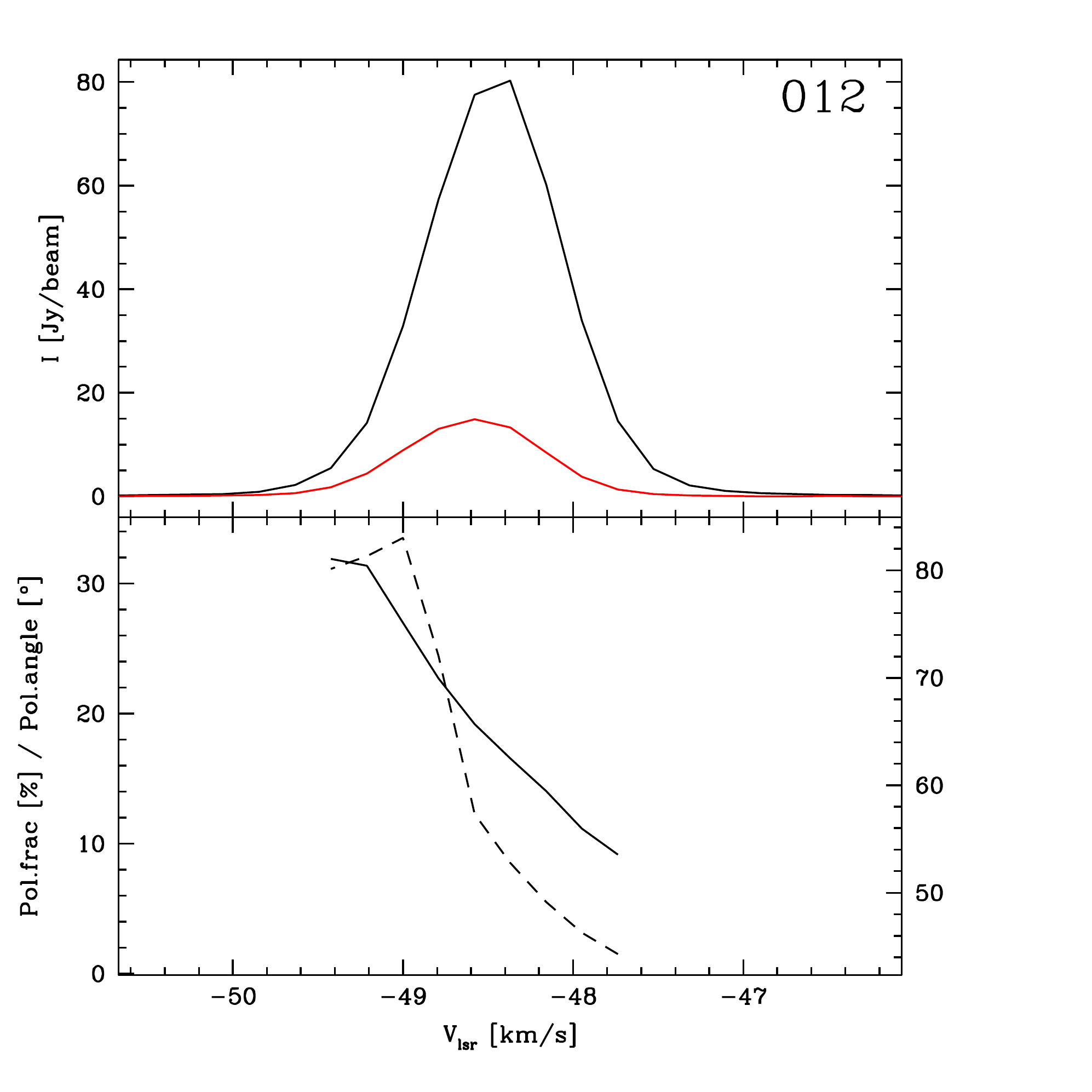}
\includegraphics[width = 6 cm]{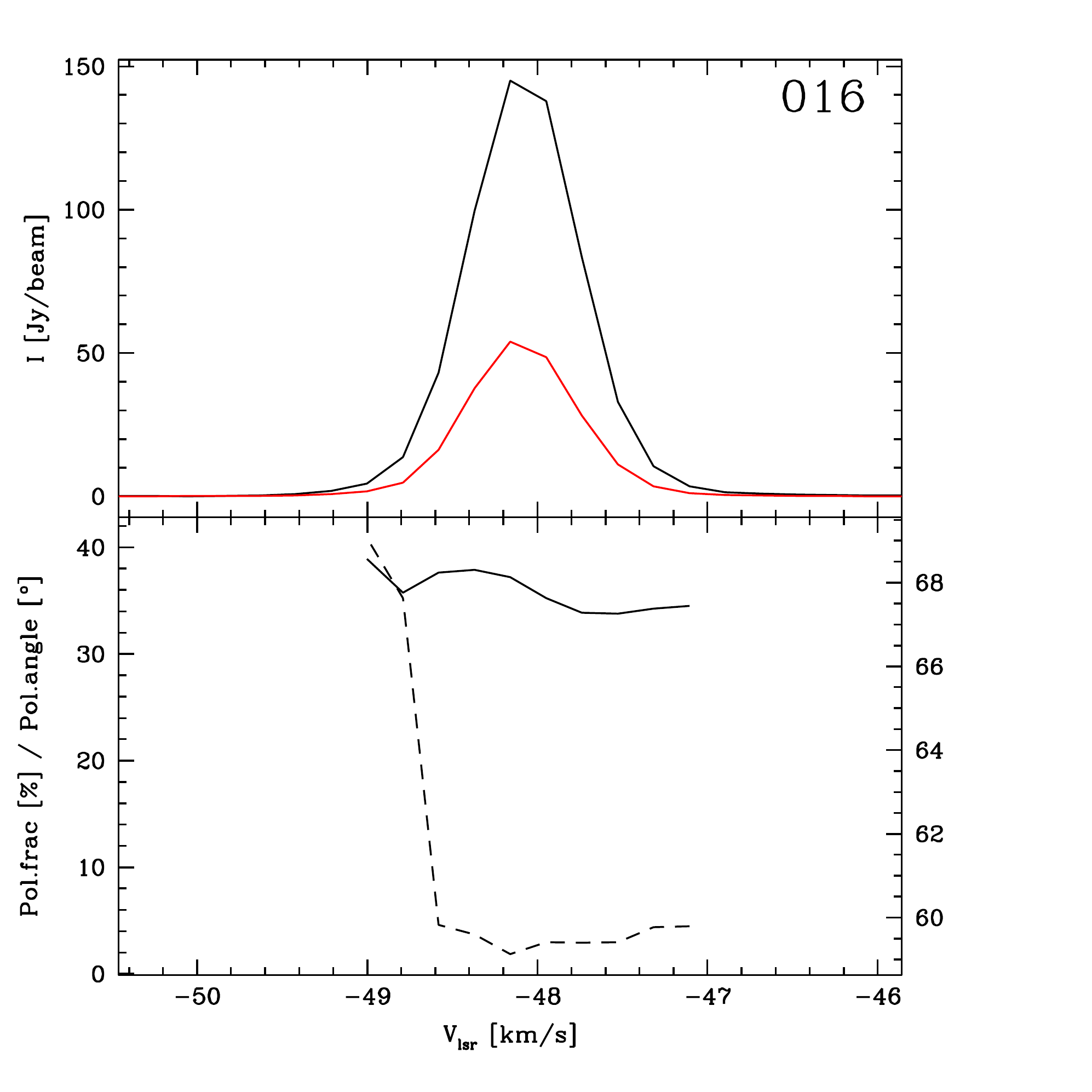}
\includegraphics[width = 6 cm]{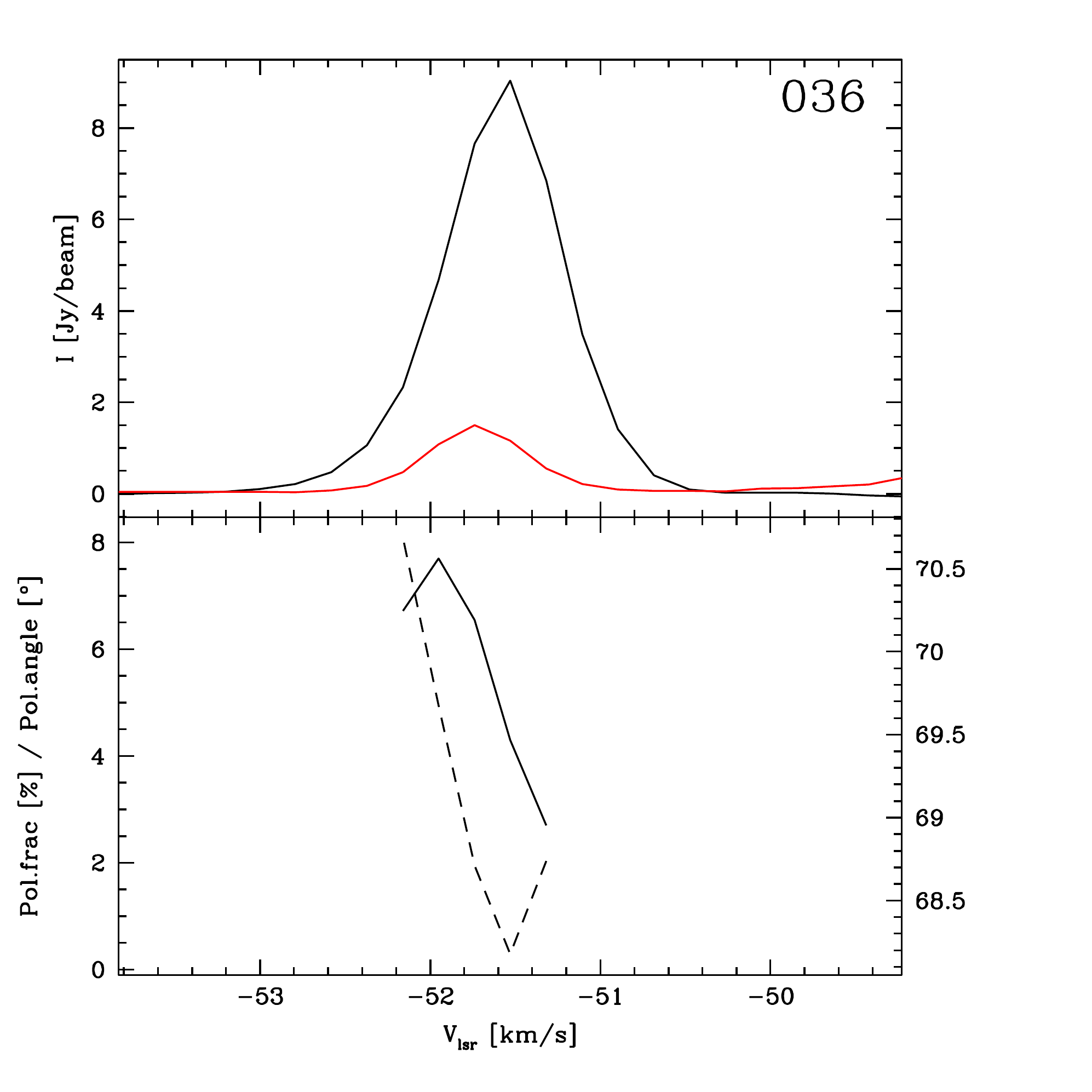}
\includegraphics[width = 6 cm]{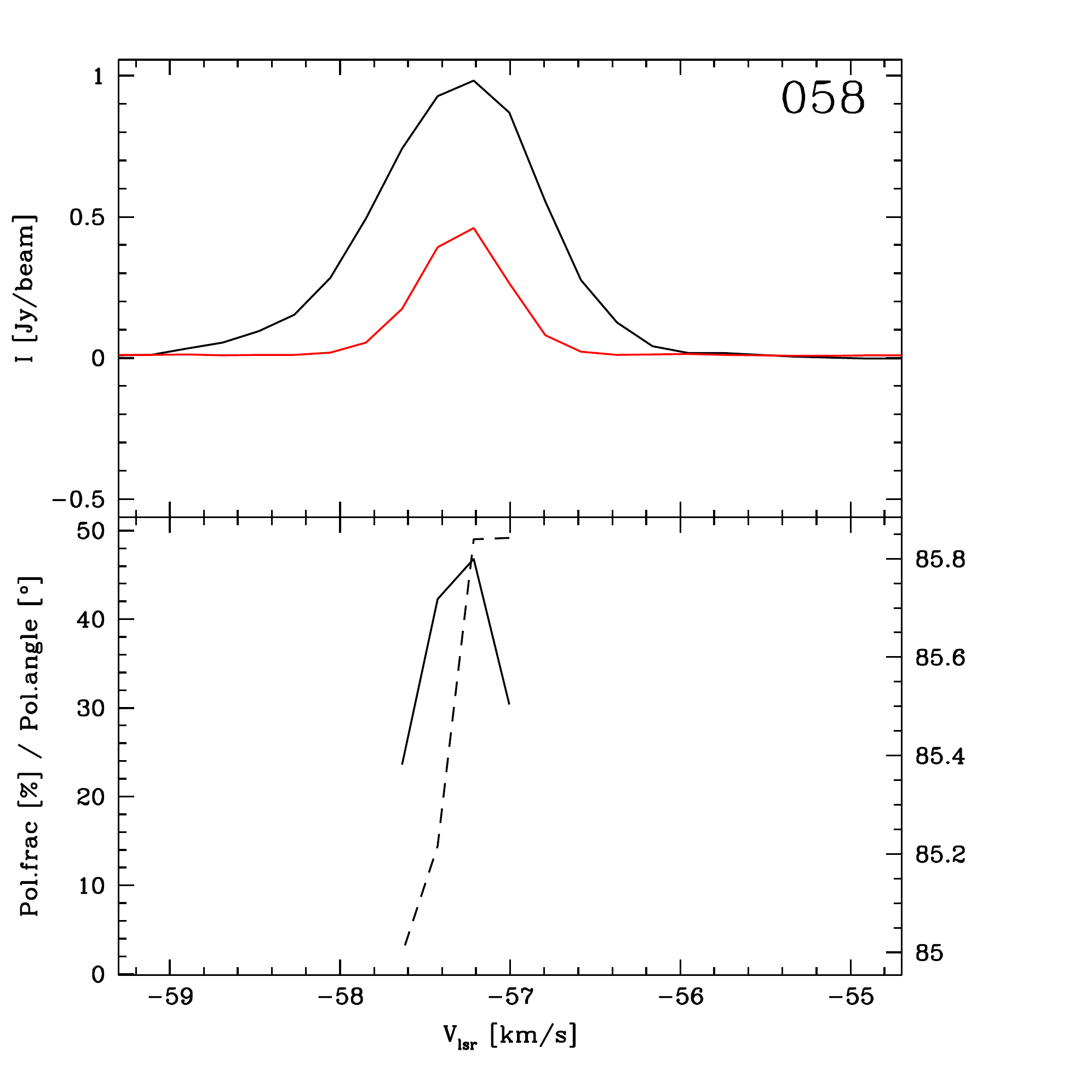}
\includegraphics[width = 6 cm]{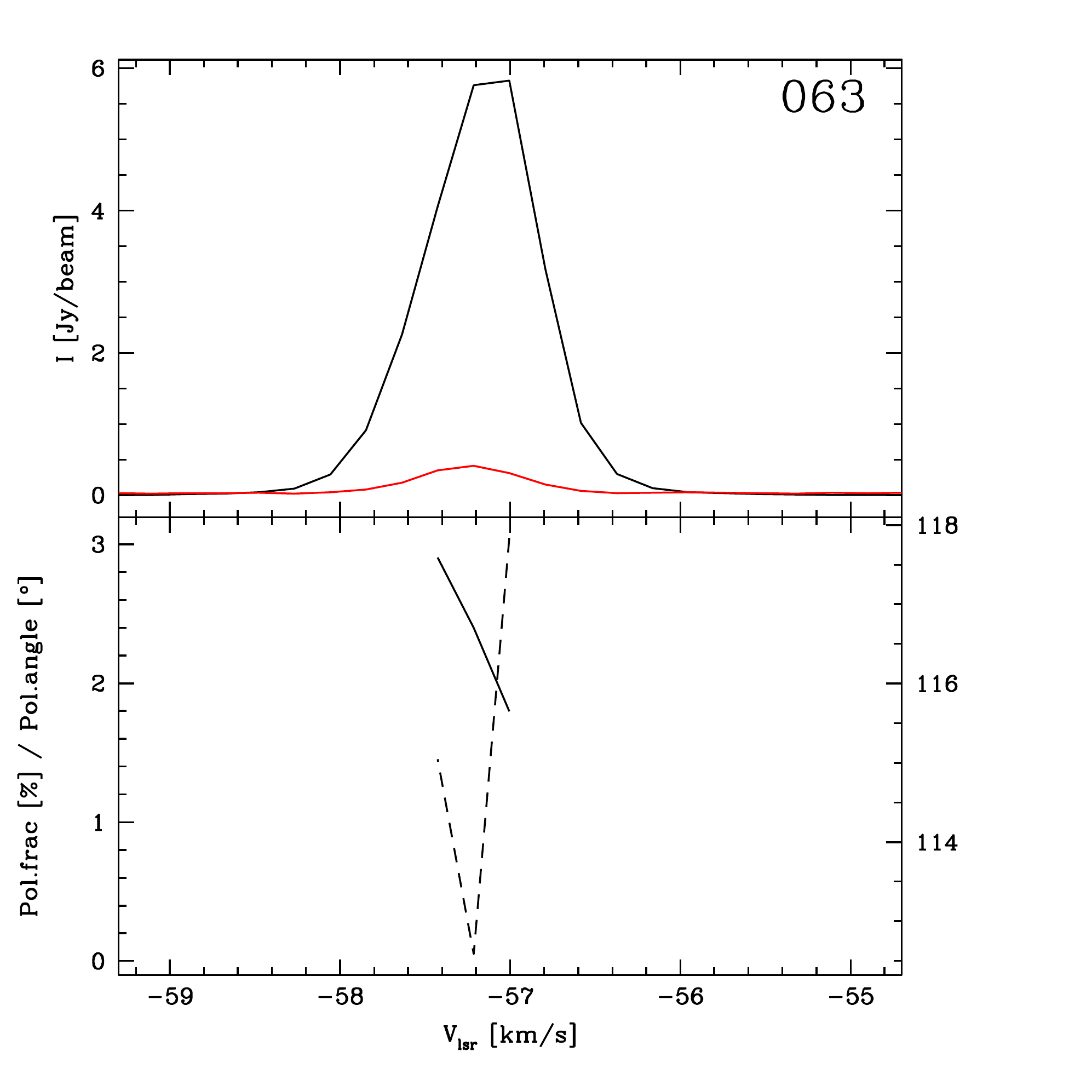}
\includegraphics[width = 6 cm]{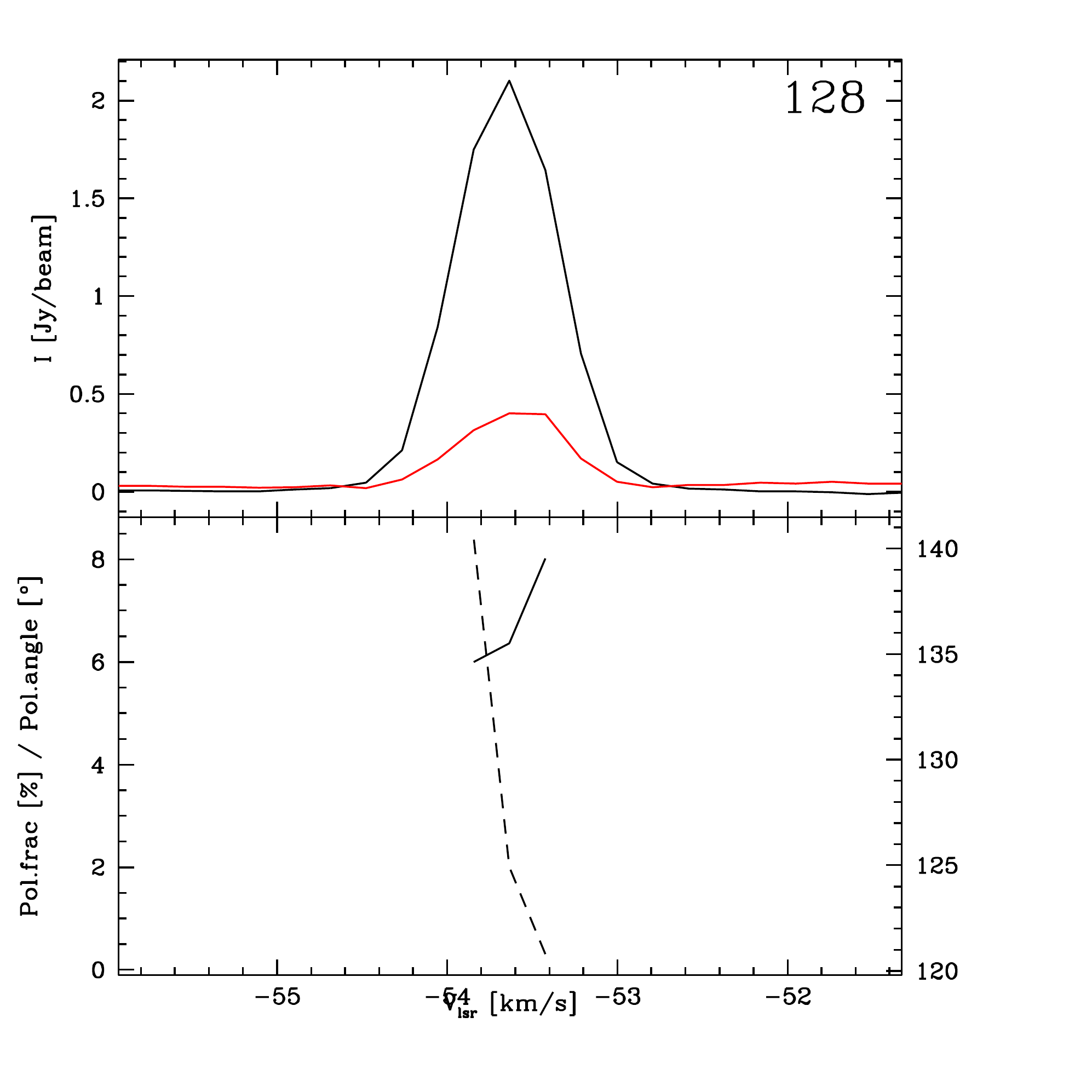}
\caption{
Total intensity ({\it I}, black solid line) and linear polarization intensity (red solid line) spectra of the \wat ~maser features 012, 016, 036, 058, 
063, and 128 ({\it upper panel}). The linear polarization intensity spectra have been multiplied by a factor of three for 036, 063, and 128. The linear 
polarization fraction (black solid line, left scale) and the linear polarization angle (dashed black line, right scale) are also shown ({\it lower panel}).
}
\label{POL_plots}
\end{figure*}

\begin{figure*} 
\centering
\includegraphics[width = 6 cm]{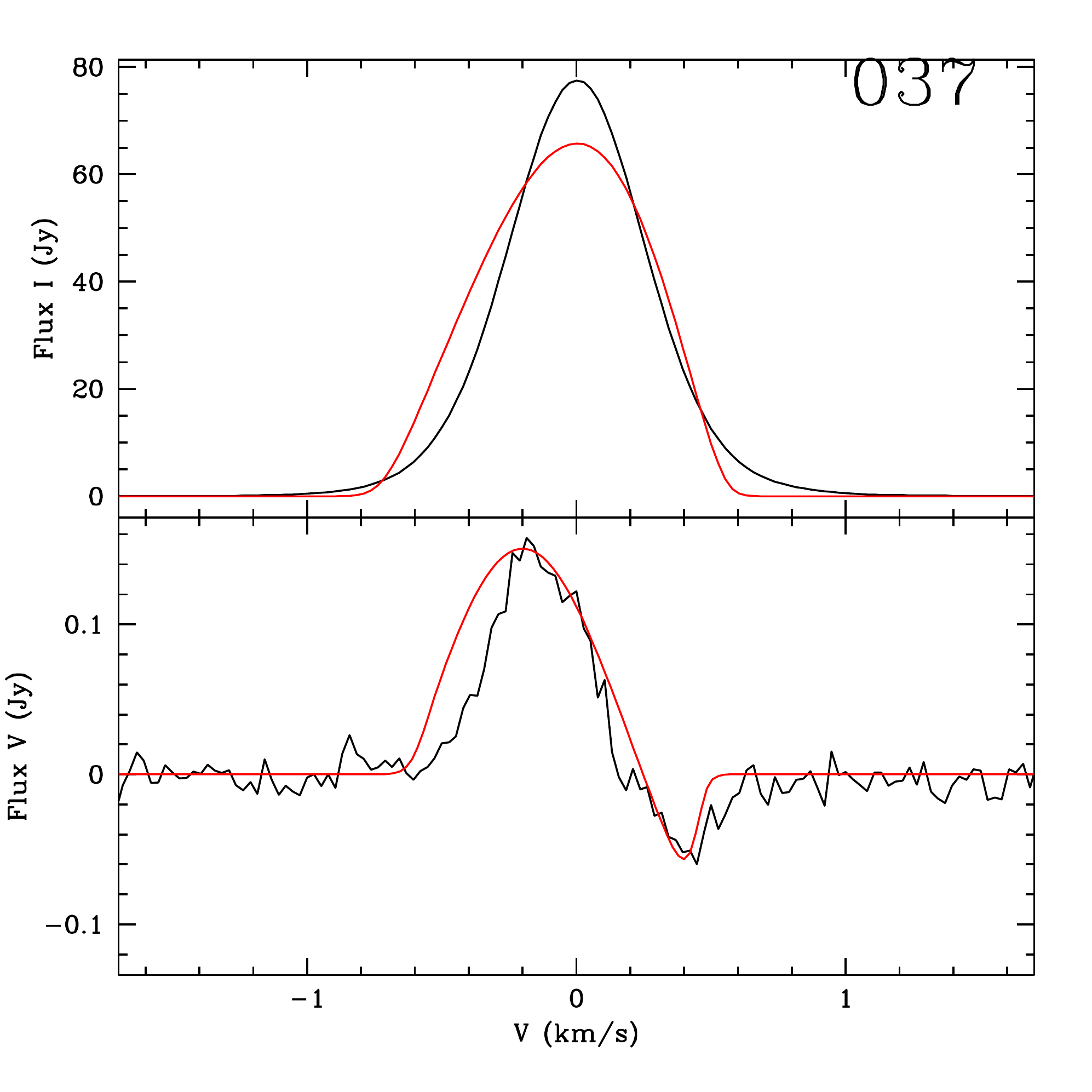}
\includegraphics[width = 6 cm]{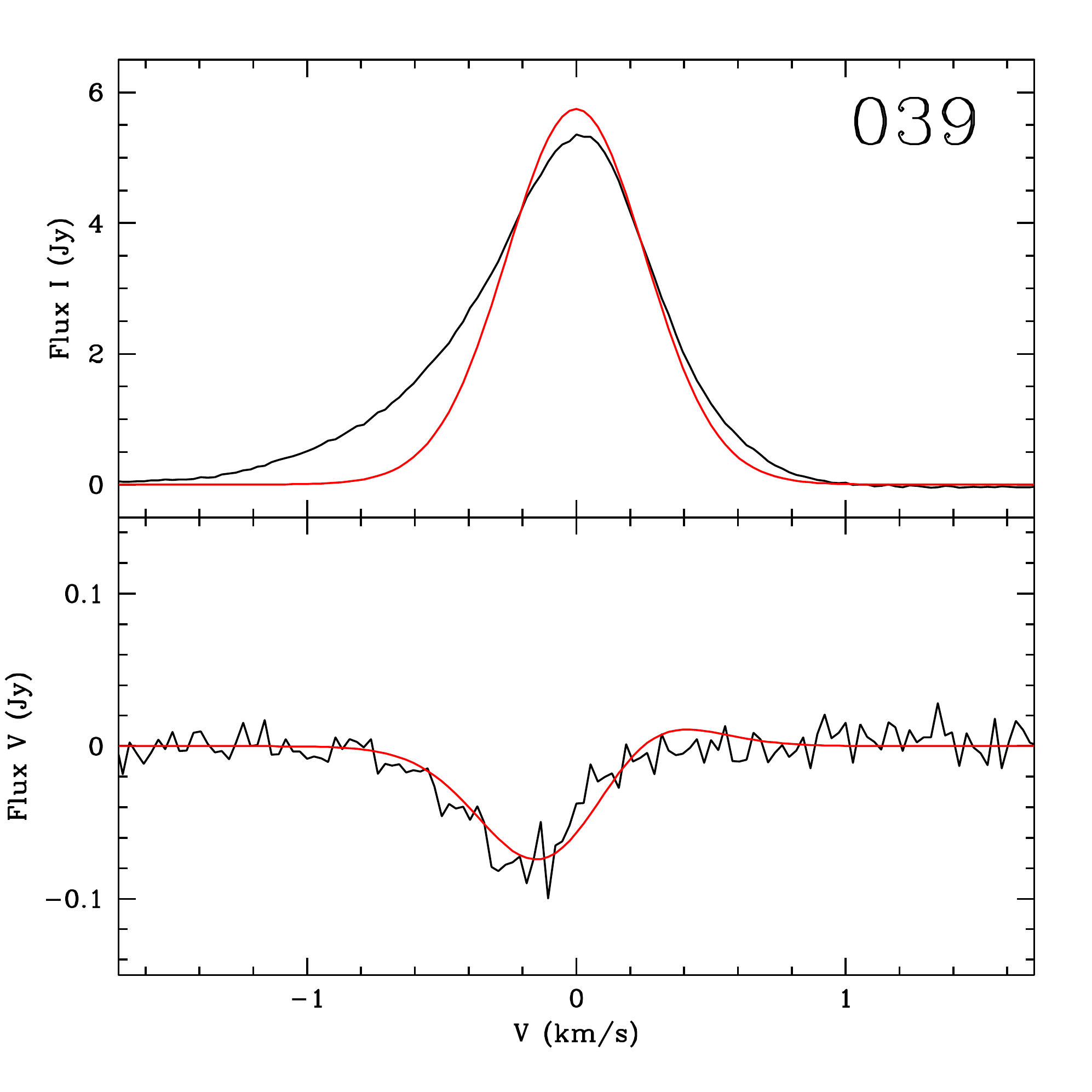}
\includegraphics[width = 6 cm]{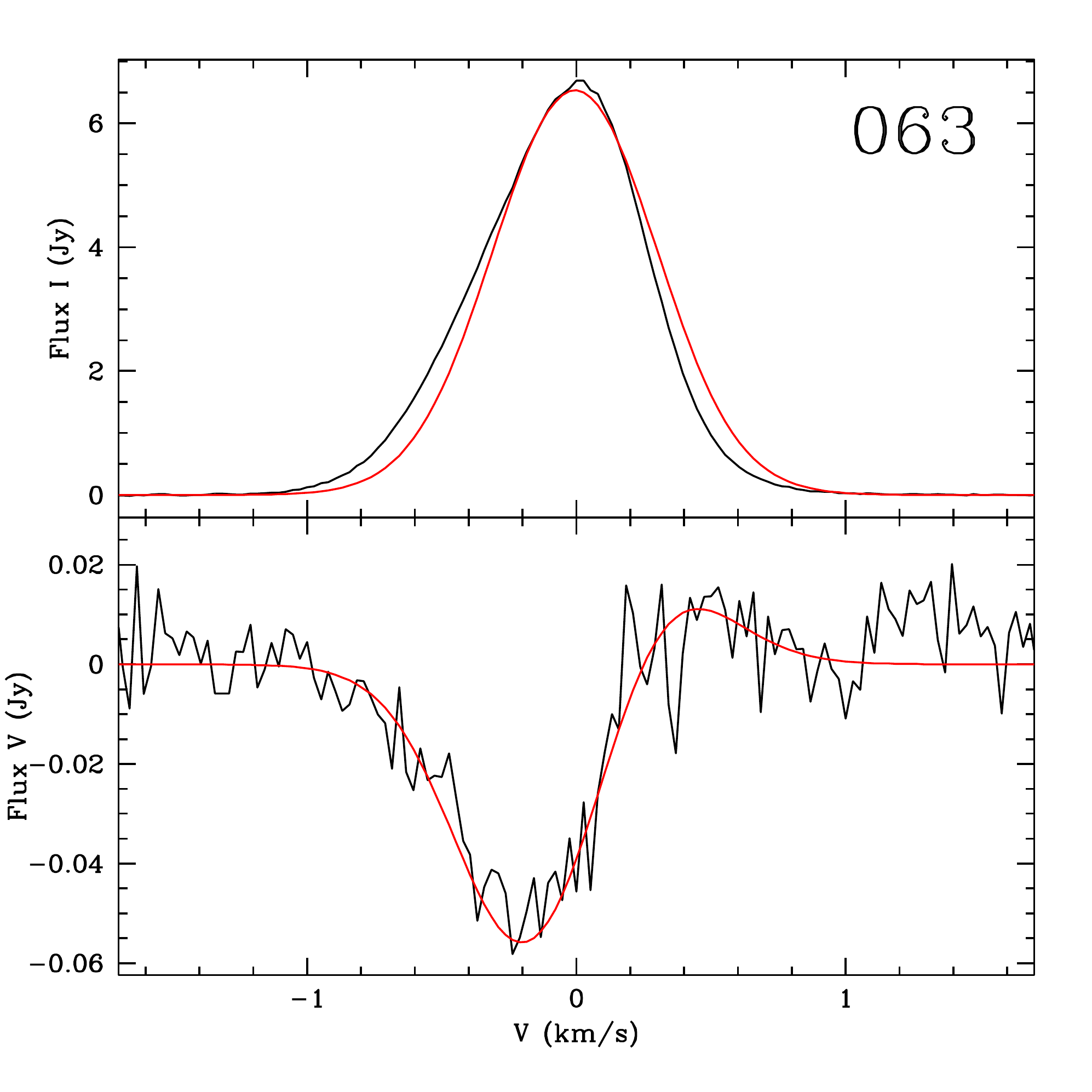}
\caption{
Total intensity ({\it I}, {\it upper panel}) and circular polarization intensity ({\it V}, {\it lower panel}) spectra for the \wat ~maser features 
037, 039, 063. The thick red line shows the best-fit models of {\it I} and {\it V} emission obtained using the FRTM code (see 
Appendix~\ref{analysis}). The maser features are centered on zero velocity.
}
\label{Vfit}
\end{figure*}
%--------------------------------------------------------------------------------------

\section{Astrometry}
\label{astro}
We did not observe in phase-referencing mode, therefore after self-calibration we cannot recover the information on the absolute position in our maser polarization maps. 
The positions  of the identified \wat\,masers  are relative to the strongest  feature used for self-calibration: ID 018 at --48.6~\kms\ and with peak flux density $\sim 2 \times 10^3$~\jyb\ (see Table~\ref{features}). 
In order to register our \wat\,maser polarization maps with previous \wat\,maser proper motion measurements (as well as with thermal emission maps), 
we used a two-steps process based on the astrometric measurements by \citet{Hachi06}. 
First, 
 we aligned the brightest features ($>$100~\jyb)
 detected in both datasets towards cluster "d" in the velocity range --48.4 to --48.8~\kms. 
These include our reference feature ID 018 and features ID 4, 5, 6, and 7 from \citet{Hachi06} (see their Table~2). 
The latter have a positional scatter  of about  10~mas in R.A. and 40~mas in DEC, around the position: $\alpha_{2000}=2^{\rm{h}}27^{\rm{m}}04^{\rm{s}}.8362$ and 
$\delta_{2000}=61^{\circ}52'24$\pas607, setting the accuracy of the registration  between the two maser datasets at the level of several tens of mas. 
In order to refine such registration to higher accuracy, we shifted  our polarized maps in such a fashion to minimise the distance among the persistent maser knots detected in the two datasets. 
For this purpose, we used only the maser cluster “a”, which displays more persistent, recognizable structures with respect to cluster “d” (affected by strong shocks) over the five years separation between the two observations. 
We found that the positional offset that maximizes the overlap among the two maser distributions is: $\Delta x = -10$~mas,  $\Delta y = +33$~mas (with respect to the position: $\alpha_{2000}=2^{\rm{h}}27^{\rm{m}}04^{\rm{s}}.8362$ and 
$\delta_{2000}=61^{\circ}52'24$\pas607). 
Incidentally, we notice that applying this offset 
brings in coincidence (within 1 mas) our strongest feature at --48.6~\kms\ (ID 018 in Table~\ref{features}) with the strongest feature detected by \citet{Hachi06} (ID 6 in  their Table~\ref{features}). 

According to this analysis, we conclude that  our absolute astrometry has an accuracy of the order of a few milliarcseconds.

\section{Geometry of the magnetic field with respect to the proper motions of water masers}
\label{B_geometry} 
According to the definition of $\theta$ as the angle between the magnetic field vector {\bf B} and the line of sight (see \S\ref{lin_pol}), we can define: 
$$ B_{\rm{l.o.s.}} = B cos(\theta), \ \ B_{\rm{sky}} = B sin(\theta) $$  
where $ B_{\rm{l.o.s.}}$ and $B_{\rm{sky}}$ are the  components of the magnetic field {\bf B} along the line-of-sight and projected in the plane of the sky, respectively. 
If now indicate with  $\phi$ the angle between $B_{\rm{sky}}$ and the proper motion orientation, we can decompose $B_{\rm{sky}}$ in the two components: 
$$B_{\rm{sky_{||}}} = B sin(\theta) cos(\phi), \ \ B_{\rm{sky_{\perp}}} = B sin(\theta) sin(\phi) $$ 
Merging the expressions above, we can write 
the components of {\bf B} parallel, $B_{\rm{||}}$, and perpendicular, $B_{\perp}$, to the proper motion orientation as follows:  
\begin{equation}
\nonumber
B_{\rm{||}} \equiv B_{\rm{sky_{||}}} = B sin(\theta) cos(\phi) 
\end{equation}
\begin{equation}
\nonumber
B_{\perp} = \sqrt{(B_{\rm{l.o.s.}})^2 + (B_{\rm{sky_{\perp}}})^2}= \sqrt{(B cos(\theta))^2 + (B sin(\theta) sin(\phi))^2} 
\end{equation}
Consequently, we can derive a ratio between these components as:  
$$ B_{\perp} / B_{\rm{||}} = \sqrt{\cot^2(\theta) / cos^2(\phi) + \tan^2(\phi)}. $$

\end{document}

%% file: table_paper_inp.tex
\begin{longtab}
  \renewcommand\thetable{1}
  \begin{longtable}{ l c c c c c c c c c c c c }
  \caption{Parameters of the 22-GHz H$_2$O maser features detected in W3(OH).} 
\scriptsize
\cr\hline
\hline
(1)&(2)   & (3)      & (4)            & (5)       & (6)              & (7)         & (8)       & (9)                     & (10)                    & (11)                        & (12)         &(13)                           \\
ID     & RA\tablefootmark{a}&Dec\tablefootmark{a}& Peak flux & $V_{\rm{lsr}}$& $\Delta v\rm{_{L}}$ &$P_{\rm{l}}$ &  $\chi$   & $\Delta V_{\rm{i}}\tablefootmark{b}$ & $T_{\rm{b}}\Delta\Omega\tablefootmark{b}$& $P_{\rm{V}}$ & $B_{\rm{l.o.s.}}$  &$\theta\tablefootmark{c}$\\
          &  offset &  offset  & Density(I)     &           &                  &             &	         &                         &                         &                &                      &      \\ 
          &(arcsec) & (arcsec) & (Jy/beam)      &  (km/s)   &      (km/s)      & (\%)        &   (\d)    & (km/s)                  & (log K sr)              &   ($\%$)       &  (mG)               &(\d)       \\ 
\hline
\endfirsthead
\caption{(continued).}\\
\hline\hline
\,\,\,\,\,(1)&(2)   & (3)      & (4)            & (5)       & (6)              & (7)         & (8)       & (9)                     & (10)                    & (11)                        & (12)         &(13)                           \\
ID     & RA\tablefootmark{a}&Dec\tablefootmark{a}& Peak flux & $V_{\rm{lsr}}$& $\Delta v\rm{_{L}}$ &$P_{\rm{l}}$ &  $\chi$   & $\Delta V_{\rm{i}}\tablefootmark{b}$ & $T_{\rm{b}}\Delta\Omega\tablefootmark{b}$& $P_{\rm{V}}$ & $B_{\rm{l.o.s.}}$  &$\theta\tablefootmark{c}$\\
          &  offset &  offset  & Density(I)     &           &                  &             &	         &                         &                         &                &                      &      \\ 
          &(arcsec) &(arcsec)  & (Jy/beam)      &  (km/s)   &      (km/s)      & (\%)        &   (\d)    & (km/s)                  & (log K sr)              &   ($\%$)       &  (mG)               &(\d)       \\ 
\hline
\endhead
\hline
\endfoot
\hline
\\
\noalign{\tablefoot{
The number ID of the detected maser features is reported in Col.~1. The positions, 
Cols.~2 and 3, refer to the brightest H$_2$O ~maser feature 018 that was used to self-calibrate the data. 
The peak flux density (I), the LSR velocity ($V_{\rm{lsr}}$), and the FWHM ($\Delta v\rm{_{L}}$) of the total intensity spectra of the 
maser features are reported in Cols.~4, 5, and 6, respectively; I, $V_{\rm{LSR}}$, and $\Delta v\rm{_{L}}$ are obtained using a Gaussian fit.
The mean linear polarization fraction, $P_{\rm{l}}$, and the mean linear polarization angles, $\chi$, are instead reported in Cols.~7 and 8, 
respectively. We determined $P_{\rm{l}}$ and $\chi$ of each H$_2$O~maser feature by  considering (more than two)  consecutive channels 
across the total intensity spectrum for which the polarized intensity is $\geq5\sigma$. 
In Col. 9 we report the intrinsic 
thermal linewidth of the maser $\Delta V_{\rm{i}}$. In Col.~10  we report the emerging brightness temperature \tbo, which is
 the product of the brightness temperature $T_{\rm{b}}$  and the solid angle of the maser beam $\Delta \Omega$.\\
\tablefoottext{a}{The reference position is assumed to be $\alpha_{2000}=2^{\rm{h}}27^{\rm{m}}04^{\rm{s}}.8362$ and 
$\delta_{2000}=61^{\circ}52'24$\pas607  (an additional offset of $\Delta x = -10$~mas,  $\Delta y = +33$~mas is also assumed in the absolute astrometry; see App.~\ref{astro}). }
\tablefoottext{b}{The best-fitting results obtained by using a model based on the radiative transfer theory of water masers 
%for $\Gamma+\Gamma_{\nu}=1~\rm{s^{-1}}$ 
\citep{Vlemmings2006a,Surcis2011a}. The errors were determined 
by analyzing the full probability distribution function.}
\tablefoottext{c}{The angle $\theta$ between the magnetic field and the maser propagation direction is determined by using the observed $P_{\rm{l}}$ 
and the fitted \tbo. The errors were determined by analyzing the full probability distribution function. We use boldface for values for which the magnetic field is more likely parallel to $\chi$.}
}}
\endlastfoot
\noalign{\bigskip}
\multicolumn{13}{c}{Cluster "d"} \\
\noalign{\smallskip}
001       & 0.0897  & -0.2034 & $0.1$& -46.26    &      $0.91$      & $-$         &  $-$      &           $-$           &           $-$           & $-$	      & $-$              	   &$-$ \\ 
002       & 0.0584  & -0.2000 & $0.06$& -46.47    &      $0.92$      & $-$         &  $-$      &           $-$           &           $-$           & $-$	      & $-$              	   &$-$ \\ 
003       & 0.0523  & -0.0694 & $0.8$& -45.63    &      $1.29$      & $-$         &  $-$      &           $-$           &           $-$           & $-$	      & $-$              	   &$-$ \\ 
004       & 0.0410  & -0.1185 & $2.9$& -46.47    &      $0.86$      & $-$         &  $-$      &           $-$           &           $-$           & $-$	      & $-$              	   &$-$ \\ 
005& 0.0397  & -0.0348 &$200.2$& -49.84   &      $0.69$      & $17.7\pm3.5$& $79\pm3$  &  $1.0^{+0.5}_{-1.4}$    & $10.8^{+0.4}_{-2.4}$& $-$       & $-$              	   &$90^{+9}_{-9}$ \\ 
006       & 0.0393  & -0.0338 & $1.7$&  -50.48   &      $0.87$      & $-$         &  $-$      &           $-$           &           $-$           & $-$	      & $-$              	   &$-$ \\ 
007       & 0.0378  & -0.0967 & $1.5$& -47.32    &      $0.87$      & $-$         &  $-$      &           $-$           &           $-$           & $-$	      & $-$              	   &$-$ \\ 
008       & 0.0276  & -0.0182 & $7.2$ & -49.52   &      $0.93$      &$42\pm19$& $-60\pm7$ &  $2.2^{+1.2}_{-0.6}$    & $6.0^{+2.1}_{-0.7}$     & $-$	      & $-$              	   &$90^{+65}_{-65}$ \\ 
009       & 0.0256  & -0.0462 & $27.3$& -48.16   &      $0.89$      & $-$         &  $-$      &           $-$           &           $-$           & $-$	      & $-$              	   &$-$ \\ 
010& 0.0247  & -0.0453 & $6.4$& -48.58    &      $1.03$      & $19.3\pm6.5$& $-72\pm5$ &  $1.3^{+0.6}_{-1.2}$    & $11.1^{+3.2}_{-0.1}$& $-$       & $-$                        &$90^{+45}_{-45}$ \\ 
011& 0.0214  & -0.1055 &$204.7$& -48.79   &      $0.98$      & $10.2\pm1.2$& $+63\pm7$ &  $1.9^{+0.7}_{-0.5}$    & $10.6^{+0.3}_{-0.3}$& $-$       & $-$              	   &$84^{+3}_{-8}$ \\ 
012& 0.0205  & -0.1066 & $80.4$& -48.37   &      $0.91$      & $19.2\pm6.5$& $+57\pm16$&  $1.1^{+0.5}_{-1.3}$    & $11.1^{+0.4}_{-2.9}$& $0.2$     & $-70\pm13$                 &$90^{+38}_{-38}$ \\ 
013       & 0.0198  & -0.1077 &$23.5$& -46.47    &      $0.83$      & $3.6\pm0.7$ & $+82\pm2$ &  $3.5^{+05}_{-1.1}$     & $9.7^{+1.1}_{-1.0}$     & $-$	      & $-$              	   &$74^{+16}_{-32}$ \\ 
014       & 0.0129  & -0.0292 & $3.1$& -47.74    &      $0.85$      & $-$         &  $-$      &           $-$           &           $-$           & $-$	      & $-$              	   &$-$ \\ 
015       & 0.0078  & -0.0455 & $5.9$& -47.95    &      $0.67$      & $-$         &  $-$      &           $-$           &           $-$           & $-$	      & $-$              	   &$-$ \\ 
016& 0.0022  & -0.0199 &$145.1$& -48.16   &      $0.75$      & $35.8\pm1.7$& $+60\pm1$ &  $1.0^{+1.2}_{-0.2}$    & $11.0^{+0.2}_{-1.0}$& $-$       & $-$              	   &$90^{+3}_{-3}$ \\ 
017       & 0.0021  & -0.0199 & $0.2$&  -50.90   &      $3.11$      & $-$         &  $-$      &           $-$           &           $-$           & $-$	      & $-$              	   &$-$ \\ 
018& 0.0000  & 0.0000 &$2050.3$&-48.58   &      $0.95$      & $15.0\pm2.7$& $+6\pm10$ &  $1.0^{+1.5}_{-0.5}$    & $11.2^{+0.2}_{-1.6}$& $0.2$      & $+59\pm9$                  &$90^{+8}_{-8}$ \\ 
019& -0.0034 & 0.0013 & $60.1$& -49.21   &      $0.72$      & $25.4\pm3.0$& $-35\pm2$ &  $1.1^{+1.6}_{-0.3}$    & $10.9^{+0.4}_{-2.4}$& $0.4$      & $-120\pm20$                &$90^{+6}_{-6}$ \\ 
020& -0.0355 & 0.0627 & $10.0$& -49.42   &      $0.54$      & $36.1\pm3.3$& $-37\pm1$ &  $1.0^{+0.8}_{-0.8}$    & $11.0^{+0.3}_{-2.7}$& $-$        & $-$              	   &$90^{+5}_{-5}$ \\ 
021& -0.1391 & -0.1918 & $34.2$& -50.48   &      $0.58$      & $15.4\pm1.0$& $-11\pm69$&  $1.0^{+0.8}_{-0.8}$    & $10.8^{+0.3}_{-0.5}$& $-$       & $-$              	   &$90^{+5}_{-5}$ \\ 
022       & -0.2376 & -0.1840 & $6.8$& -47.95    &      $1.43$      & $-$         &  $-$      &           $-$           &           $-$           & $-$	      & $-$              	   &$-$ \\ 
023       & -0.2407 & -0.1861 & $0.3$& -47.11    &      $0.81$      & $-$         &  $-$      &           $-$           &           $-$           & $-$	      & $-$              	   &$-$ \\ 
024       & -0.2489 & -0.2023 & $1.0$&  -59.95   &      $0.73$      & $-$         &  $-$      &           $-$           &           $-$           & $-$	      & $-$              	   &$-$ \\ 
025       & -0.2590 & -0.2059 & $1.6$&  -60.16   &      $0.81$      & $-$         &  $-$      &           $-$           &           $-$           & $-$	      & $-$              	   &$-$ \\ 
026       & -0.2598 & -0.2051 & $0.2$&  -59.11   &      $0.61$      & $-$         &  $-$      &           $-$           &           $-$           & $-$	      & $-$              	   &$-$ \\ 
027       & -0.2615 & -0.2054 & $0.3$&  -59.74   &      $0.76$      & $-$         &  $-$      &           $-$           &           $-$           & $-$	      & $-$              	   &$-$ \\ 
028       & -0.2625 & -0.2049 & $0.08$&  -58.90   &      $0.70$      & $-$         &  $-$      &           $-$           &           $-$           & $-$	      & $-$              	   &$-$ \\ 
\noalign{\bigskip}
\multicolumn{13}{c}{Cluster "a"} \\
\noalign{\smallskip}
029       & -0.6072 & 0.0486  & $1.8$& -43.95    &    $0.53$        & $-$         & $-$       & $-$                     & $-$                     & $-$	      & $-$              	   &$-$\\ 
030       & -0.6344 & 0.0427  & $0.1$& -54.06    &    $1.13$        & $-$         & $-$       & $-$                     & $-$                     & $-$	      & $-$              	   &$-$\\ 
031       & -0.6354 & 0.0438  & $5.8$& -58.90    &    $0.89$        & $-$         & $-$       & $-$                     & $-$                     & $-$	      & $-$              	   &$-$\\ 
032       & -0.6358 & 0.0461  & $9.2$& -56.16    &    $0.71$        & $1.7\pm0.2$ & $+62\pm3$ & $3.7^{+0.4}_{-0.8}$     & $9.3^{+1.2}_{-1.6}$     & $-$	      & $-$              	   &$\mathbf{62^{+5}_{-47}}$ \\ 
033       & -0.6366 & 0.0439  & $3.4$& -58.90    &    $0.82$        & $-$         & $-$       & $-$                     & $-$                     & $-$	      & $-$              	   &$-$\\ 
034       & -0.6371 & 0.0439  & $7.0$& -59.32    &    $0.79$        & $-$         & $-$       & $-$                     & $-$                     & $-$	      & $-$              	   &$-$\\ 
035       & -0.6380 & 0.0438  & $1.9$& -59.32    &    $0.73$        & $-$         & $-$       & $-$                     & $-$                     & $-$	      & $-$              	   &$-$\\ 
036       & -0.6416 & 0.0537  & $9.0$& -51.53    &    $0.76$        & $6.6\pm1.2$ & $+69\pm1$ & $3.5^{+0.5}_{-0.9}$     & $8.7^{+3.1}_{-0.3}$     & $0.8$	      & $-356\pm107$               &$90^{+18}_{-18}$ \\ 
037& -0.6425 & 0.0542 &$71.8$& -51.74&$0.61$       & $1.4\pm0.1$ & $+69\pm1$ & $1.0^{+0.8}_{-0.1}$     & $10.7^{+0.2}_{-0.6}$    & $0.3$	      & $-60\pm11$                 &$\mathbf{51^{+1}_{-37}}$ \\ 
038       & -0.6428 & 0.0476  & $1.6$& -60.58    &    $0.87$        & $-$         & $-$       & $-$                     & $-$                     & $-$	      & $-$              	   &$-$\\ 
039       & -0.6433 & 0.0546  & $4.8$& -51.53    &    $0.68$        & $11.8\pm3.0$ & $+88\pm1$& $2.5^{+0.7}_{-1.0}$     & $8.1^{+0.5}_{-2.0}$     & $1.6$	      & $+411\pm114$               &$90^{+14}_{-14}$ \\ 
040       & -0.6437 & 0.0498  & $0.7$& -55.32    &    $0.77$        & $-$         & $-$       & $-$                     & $-$                     & $-$	      & $-$              	   &$-$\\ 
041       & -0.6453 & 0.0501  & $0.7$& -55.74    &    $0.70$        & $-$         & $-$       & $-$                     & $-$                     & $-$	      & $-$              	   &$-$\\ 
042       & -0.6515 & 0.0535  & $0.2$& -54.27    &    $0.84$        & $-$         & $-$       & $-$                     & $-$                     & $-$	      & $-$              	   &$-$\\ 
043& -0.6520 & 0.0590 &$28.0$& -51.53    & $0.64$  & $0.9\pm0.1$ & $+69\pm4$ & $1.6^{+0.5}_{-0.7}$     & $10.1^{+0.7}_{-1.0}$    & $-$	      & $-$              	   &$\mathbf{60^{+2}_{-46}}$ \\ 
044       & -0.6622 & 0.0630  & $2.3$& -52.16    &    $0.53$        & $-$         & $-$       & $-$                     & $-$                     & $-$	      & $-$              	   &$-$\\ 
045       & -0.6633 & 0.0633  & $2.4$& -52.58    &    $0.53$        & $-$         & $-$       & $-$                     & $-$                     & $-$	      & $-$              	   &$-$\\ 
046       & -0.6643 & 0.0624  & $0.9$& -53.42    &    $0.56$        & $-$         & $-$       & $-$                     & $-$                     & $-$	      & $-$              	   &$-$\\ 
047       & -0.6650 & 0.0621  & $2.2$& -53.21    &    $1.06$        & $-$         & $-$       & $-$                     & $-$                     & $-$	      & $-$              	   &$-$\\ 
048       & -0.6650 & 0.0636  & $3.9$& -53.00    &    $0.50$        & $-$         & $-$       & $-$                     & $-$                     & $-$	      & $-$              	   &$-$\\ 
049       & -0.666 & 0.0624  & $5.3$& -53.63    &    $0.56$        & $-$         & $-$       & $-$                     & $-$                     & $-$	      & $-$              	   &$-$\\ 
050& -0.6665 & 0.0647 &$11.5$& -53.00 & $0.56$     & $1.2\pm0.1$ & $+65\pm1$ & $1.2^{+0.3}_{-0.1}$     & $10.4^{+0.2}_{-0.3}$    & $-$	      & $-$              	   &$\mathbf{51^{+1}_{-35}}$ \\ 
051& -0.6667 & 0.0648 &$11.0$& -53.00 & $0.54$     & $1.3\pm0.1$ & $+63\pm1$ & $1.2^{+0.5}_{-0.2}$     & $10.3^{+0.3}_{-0.3}$    & $0.4$	      & $-84\pm35$                 &$\mathbf{49^{+1}_{-32}}$ \\ 
052       & -0.6676 & 0.0649  & $4.1$& -53.00    &    $0.67$        & $-$         & $-$       & $-$                     & $-$                     & $-$	      & $-$              	   &$-$\\ 
053       & -0.7339 & -0.1245 & $0.5$& -50.69    &    $0.88$        & $-$         & $-$       & $-$                     & $-$                     & $-$	      & $-$              	   &$-$\\ 
054       & -0.7948 &-0.1137  & $0.3$& -58.69    &    $0.70$        & $-$         & $-$       & $-$                     & $-$                     & $-$	      & $-$              	   &$-$\\ 
055       & -0.7976 &-0.1132  & $0.6$& -58.48    &    $0.74$        & $-$         & $-$       & $-$                     & $-$                     & $-$	      & $-$              	   &$-$\\ 
056       & -0.8079 &-0.0662  &$12.0$& -60.58    &    $0.84$        & $-$         & $-$       & $-$                     & $-$                     & $-$	      & $-$              	   &$-$\\ 
057& -0.8112 & -0.0680&$60.9$& -57.22    & $0.61$  & $6.0\pm0.5$ & $-81\pm46$& $1.4^{+0.6}_{-0.9}$     & $10.3^{+0.9}_{-1.9}$    & $-$	      & $-$              	   &$79^{+7}_{-7}$ \\ 
058       & -0.8128 & -0.0679 & $1.0$& -57.21    &    $0.98$        & $42.3\pm8.2$& $+86\pm1$ & $1.7^{+1.1}_{-0.5}$     & $8.3^{+0.5}_{-2.0}$     & $-$	      & $-$              	   &$90^{+47}_{-47}$ \\ 
059       & -0.8129 &-0.0678  & $1.0$& -57.22    &    $0.98$        & $-$         & $-$       & $-$                     & $-$                     & $-$	      & $-$              	   &$-$\\ 
060       & -0.8169 & -0.0628 & $1.8$& -48.37    &    $1.20$        & $22.3\pm4.0$ & $+63\pm3$& $1.5^{+1.5}_{-0.3}$     & $8.4^{+2.6}_{-0.1}$     & $-$	      & $-$              	   &$90^{+8}_{-8}$ \\ 
061       & -0.8170 & -0.0629 & $1.3$& -49.63    &    $2.12$        & $-$         & $-$       & $-$                     & $-$                     & $-$	      & $-$              	   &$-$\\ 
062       & -0.8172 & -0.0692 & $1.3$& -57.00    &    $0.57$        & $-$         & $-$       & $-$                     & $-$                     & $-$	      & $-$              	   &$-$\\ 
063       & -0.8182 & -0.0701 & $5.8$& -57.00    &    $0.61$        & $2.4\pm0.6$ & $-65\pm3$ & $3.0^{+0.6}_{-0.7}$     & $8.8^{+0.7}_{-1.9}$     & $0.9$	      & $+360\pm120$               &$90^{+48}_{-48}$ \\ 
064       & -0.8199 & -0.0643 & $1.7$& -56.58    &    $1.33$        & $-$         & $-$       & $-$                     & $-$                     & $-$	      & $-$              	   &$-$\\ 
065       & -0.8206 & -0.0643 & $1.0$& -49.84    &    $3.31$        & $-$         & $-$       & $-$                     & $-$                     & $-$	      & $-$              	   &$-$\\ 
066       & -0.8213 & -0.0644 & $1.1$& -47.32    &    $1.86$        & $-$         & $-$       & $-$                     & $-$                     & $-$	      & $-$              	   &$-$\\ 
067       & -0.8227 & -0.0011 & $1.7$& -56.37    &    $0.63$        & $-$         & $-$       & $-$                     & $-$                     & $-$	      & $-$              	   &$-$\\ 
068       & -0.8247 & -0.0706 & $0.4$& -56.79    &    $0.56$        & $-$         & $-$       & $-$                     & $-$                     & $-$	      & $-$              	   &$-$\\ 
069& -0.8259 & -0.0003& $4.0$& -50.27    & $1.05$  & $5.0\pm0.1$ & $-66\pm2$ & $1.3^{+0.4}_{-1.4}$     & $11.0^{+0.6}_{-1.1}$    & $-$	      & $-$              	   &$67^{+15}_{-2}$ \\ 
070       & -0.8261 & -0.0706 & $0.8$& -55.95    &    $1.24$        & $-$         & $-$       & $-$                     & $-$                     & $-$	      & $-$              	   &$-$\\ 
071       & -0.8275 & -0.0689 & $1.8$& -52.37    &    $0.71$        & $-$         & $-$       & $-$                     & $-$                     & $-$	      & $-$              	   &$-$\\ 
072       & -0.8278 & -0.0714 & $0.8$& -55.95    &    $0.74$        & $-$         & $-$       & $-$                     & $-$                     & $-$	      & $-$              	   &$-$\\ 
073       & -0.8284 & -0.0713 & $1.5$& -55.95    &    $0.75$        & $-$         & $-$       & $-$                     & $-$                     & $-$	      & $-$              	   &$-$\\ 
074       & -0.8323 & -0.0709 & $0.08$& -54.69    &    $0.73$        & $-$         & $-$       & $-$                     & $-$                     & $-$	      & $-$              	   &$-$\\ 
075       & -0.8370 & -0.0709 & $0.2$& -54.90    &    $1.14$        & $-$         & $-$       & $-$                     & $-$                     & $-$	      & $-$              	   &$-$\\ 
076       & -0.8405 & -0.0019 & $0.8$& -43.53    &    $0.68$        & $-$         & $-$       & $-$                     & $-$                     & $-$	      & $-$              	   &$-$\\ 
077       & -0.8435 & -0.0021 & $1.2$& -56.58    &    $0.67$        & $-$         & $-$       & $-$                     & $-$                     & $-$	      & $-$              	   &$-$\\ 
078       & -0.8440 & -0.0015 & $0.6$& -55.53    &    $1.10$        & $-$         & $-$       & $-$                     & $-$                     & $-$	      & $-$              	   &$-$\\ 
079       & -0.8443 & -0.0025 & $3.6$& -56.58    &    $0.93$        & $-$         & $-$       & $-$                     & $-$                     & $-$	      & $-$              	   &$-$\\ 
080       & -0.8450 &-0.0040  & $0.9$& -58.69    &    $0.70$        & $-$         & $-$       & $-$                     & $-$                     & $-$	      & $-$              	   &$-$\\ 
081       & -0.8458 & -0.0756 & $0.7$& -55.32    &    $0.74$        & $-$         & $-$       & $-$                     & $-$                     & $-$	      & $-$              	   &$-$\\ 
082       & -0.8474 &-0.0754  & $0.09$& -59.11    &    $0.83$        & $-$         & $-$       & $-$                     & $-$                     & $-$	      & $-$              	   &$-$\\ 
083       & -0.8499 &-0.0748  & $0.2$& -59.74    &    $0.65$        & $-$         & $-$       & $-$                     & $-$                     & $-$	      & $-$              	   &$-$\\ 
084       & -0.8684 &-0.0666  & $1.6$& -60.58    &    $1.16$        & $-$         & $-$       & $-$                     & $-$                     & $-$	      & $-$              	   &$-$\\ 
085       & -0.9286 & -0.0048 & $0.07$& -55.74    &    $0.68$        & $-$         & $-$       & $-$                     & $-$                     & $-$	      & $-$              	   &$-$\\ 
086       & -0.9455 & -0.0023 & $1.6$& -56.79    &    $0.70$        & $-$         & $-$       & $-$                     & $-$                     & $-$	      & $-$              	   &$-$\\ 
087       & -0.9481 & -0.0027 & $0.5$& -57.85    &    $1.40$        & $-$         & $-$       & $-$                     & $-$                     & $-$	      & $-$              	   &$-$\\ 
088       & -0.9491 & 0.0122  & $0.3$& -55.11    &    $0.70$        & $-$         & $-$       & $-$                     & $-$                     & $-$	      & $-$              	   &$-$\\ 
089       & -0.9501 & 0.0130  & $3.1$& -53.85    &    $0.63$        & $-$         & $-$       & $-$                     & $-$                     & $-$	      & $-$              	   &$-$\\ 
090       & -0.9513 & 0.0132  & $0.5$& -54.48    &    $0.66$        & $-$         & $-$       & $-$                     & $-$                     & $-$	      & $-$              	   &$-$\\ 
091       & -0.9540 & 0.0152  & $6.5$& -50.48    &    $0.69$        & $-$         & $-$       & $-$                     & $-$                     & $-$	      & $-$              	   &$-$\\ 
092& -0.9560 & 0.0151&$15.7$& -53.85    & $1.90$   & $1.4\pm0.1$ & $+69\pm2$ & $2.1^{+0.1}_{-1.9}$     & $10.8^{+0.2}_{-1.8}$    & $-$	      & $-$              	   &$\mathbf{59^{+1}_{-46}}$ \\ 
093       & -0.9563 & 0.0154  & $7.6$& -52.37    &    $2.00$        & $-$         & $-$       & $-$                     & $-$                     & $-$	      & $-$              	   &$-$\\ 
094       & -0.9565 & -0.0485 & $0.2$& -55.53    &    $0.81$        & $-$         & $-$       & $-$                     & $-$                     & $-$	      & $-$              	   &$-$\\ 
095       & -0.9568 &-0.0484  & $1.1$& -57.85    &    $0.95$        & $-$         & $-$       & $-$                     & $-$                     & $-$	      & $-$              	   &$-$\\ 
096       & -0.9571 & -0.0479 & $0.9$& -57.85    &    $0.63$        & $-$         & $-$       & $-$                     & $-$                     & $-$	      & $-$              	   &$-$\\ 
097       & -0.9574 & -0.0480 & $0.04$& -55.74    &    $0.52$        & $-$         & $-$       & $-$                     & $-$                     & $-$	      & $-$              	   &$-$\\ 
098       & -0.9578 & 0.0139  & $2.4$& -43.10    &    $0.80$        & $-$         & $-$       & $-$                     & $-$                     & $-$	      & $-$              	   &$-$\\ 
099       & -0.9582 & 0.0140  & $2.5$& -43.32    &    $0.92$        & $-$         & $-$       & $-$                     & $-$                     & $-$	      & $-$              	   &$-$\\ 
100       & -0.9591 & 0.0141  & $3.1$& -43.74    &    $0.90$        & $-$         & $-$       & $-$                     & $-$                     & $-$	      & $-$              	   &$-$\\ 
101       & -0.9595 & 0.0142  & $4.7$& -44.79    &    $1.08$        & $-$         & $-$       & $-$                     & $-$                     & $-$	      & $-$              	   &$-$\\ 
102       & -0.9596 & 0.0135  & $1.2$& -58.06    &    $1.02$        & $-$         & $-$       & $-$                     & $-$                     & $-$	      & $-$              	   &$-$\\ 
103       & -0.9603 & 0.0162  & $0.6$& -51.32    &    $0.58$        & $-$         & $-$       & $-$                     & $-$                     & $-$	      & $-$              	   &$-$\\ 
104& -0.9606 & 0.0145&$24.2$& -45.21    & $0.76$   & $0.9\pm0.1$ & $+69\pm1$ & $1.3^{+1.5}_{-0.3}$     & $10.6^{+0.1}_{-0.3}$    & $0.6$	      & $+169\pm34$                &$\mathbf{55^{+3}_{-41}}$ \\ 
105& -0.9611 & 0.0156&$17.2$& -53.85    & $0.81$   & $1.5\pm0.1$ & $+86\pm7$ & $1.3^{+1.5}_{-0.2}$     & $10.7^{+0.4}_{-0.9}$    & $0.5$	      & $+161\pm41$                &$\mathbf{60^{+1}_{-45}}$ \\ 
106       & -0.9615 & 0.0149  & $4.5$& -45.63    &    $0.78$        & $-$         & $-$       & $-$                     & $-$                     & $-$	      & $-$              	   &$-$\\ 
107       & -0.9620 & 0.0167  & $5.5$& -51.74    &    $0.77$        & $-$         & $-$       & $-$                     & $-$                     & $-$	      & $-$              	   &$-$\\ 
108       & -0.9627 & 0.0157  & $0.2$& -46.90    &    $0.65$        & $-$         & $-$       & $-$                     & $-$                     & $-$	      & $-$              	   &$-$\\ 
109       & -0.9628 & 0.0158  & $7.7$& -49.84    &    $0.82$        & $-$         & $-$       & $-$                     & $-$                     & $-$	      & $-$              	   &$-$\\ 
110       & -0.9632 & 0.0139  & $0.3$& -42.89    &    $0.95$        & $-$         & $-$       & $-$                     & $-$                     & $-$	      & $-$              	   &$-$\\ 
111       & -0.9632 & 0.0168  & $2.6$& -51.32    &    $1.00$        & $-$         & $-$       & $-$                     & $-$                     & $-$	      & $-$              	   &$-$\\ 
112       & -0.9701 & 0.0137  & $5.7$& -59.32    &    $1.61$        & $-$         & $-$       & $-$                     & $-$                     & $-$	      & $-$              	   &$-$\\ 
113       & -0.9703 & 0.0141  & $0.1$& -56.58    &    $2.00$        & $-$         & $-$       & $-$                     & $-$                     & $-$	      & $-$              	   &$-$\\ 
114       & -0.9726 & 0.0516  & $0.03$& -39.74    &    $0.94$        & $-$         & $-$       & $-$                     & $-$                     & $-$	      & $-$              	   &$-$\\ 
115       & -0.9738 & 0.0125  & $0.5$& -51.32    &    $0.65$        & $-$         & $-$       & $-$                     & $-$                     & $-$	      & $-$              	   &$-$\\ 
116       & -0.9746 & 0.0136  & $0.8$& -60.16    &    $2.13$        & $-$         & $-$       & $-$                     & $-$                     & $-$	      & $-$              	   &$-$\\ 
117       & -0.9815 &-0.0441  & $0.8$& -60.16    &    $0.64$        & $-$         & $-$       & $-$                     & $-$                     & $-$	      & $-$              	   &$-$\\ 
118       & -0.9816 &-0.0441  & $0.7$& -60.16    &    $0.62$        & $-$         & $-$       & $-$                     & $-$                     & $-$	      & $-$              	   &$-$\\ 
119       & -0.9849 &-0.0415  & $0.2$& -58.48    &    $1.12$        & $-$         & $-$       & $-$                     & $-$                     & $-$	      & $-$              	   &$-$\\ 
\noalign{\bigskip}
\multicolumn{13}{c}{Cluster "e"} \\
\noalign{\smallskip}
120       & -1.1899 & -0.3140 & $0.5$& -50.90    &    $0.49$        & $-$         & $-$       & $-$                     & $-$                     & $-$	      & $-$              	   &$-$\\ 
121& -1.1933 & -0.3141& $7.8$& -50.90 & $0.60$     & $2.1\pm0.1$ & $+52\pm1$ & $1.1^{+0.5}_{-0.1}$     & $10.5^{+0.2}_{-0.4}$    & $-$	      & $-$              	   &$\mathbf{62^{+15}_{-41}}$ \\ 
122       & -1.2056 & -0.2854 & $0.4$& -51.32    &    $0.51$        & $-$         & $-$       & $-$                     & $-$                     & $-$	      & $-$              	   &$-$\\ 
\noalign{\bigskip}
\multicolumn{13}{c}{Cluster "b"} \\
\noalign{\smallskip}
123       & -1.4045 & 0.1428  & $5.5$& -51.11    &    $0.48$        & $8.0\pm1.5$ & $+37\pm7$ & $1.6^{+0.7}_{-0.4}$     & $9.5^{+0.7}_{-1.8}$     & $-$	      & $-$              	   &$90^{+11}_{-11}$ \\ 
124       & -1.4384 & 0.1261  & $1.8$& -46.90    &    $0.58$        & $-$         & $-$       & $-$                     & $-$                     & $-$	      & $-$              	   &$-$\\ 
125       & -1.4432 & 0.1242  & $5.6$& -44.16    &    $0.66$        & $-$         & $-$       & $-$                     & $-$                     & $-$	      & $-$              	   &$-$\\ 
126       & -1.4908 & 0.1022  & $0.3$& -48.79    &    $1.64$        & $-$         & $-$       & $-$                     & $-$                     & $-$	      & $-$              	   &$-$\\ 
127       & -1.5437 & 0.0778  & $0.3$& -51.11    &    $0.67$        & $-$         & $-$       & $-$                     & $-$                     & $-$	      & $-$              	   &$-$\\   
128& -1.7615 & 0.1491& $2.1$& -53.63&$0.69$        & $6.4\pm1.0$ & $-55\pm10$&  $1.0^{+0.7}_{-0.8}$    & $10.3^{+0.7}_{-2.2}$    & $-$	      & $-$              	   &$80^{+8}_{-7}$ \\ 
129       & -1.7825 & 0.1442  & $0.2$& -54.27    &      $0.59$      & $-$         & $-$       &  $-$             & $-$              & $-$	      & $-$              	   &$-$ \\ 
\noalign{\bigskip}
\multicolumn{13}{c}{Cluster "c"} \\
\noalign{\smallskip}
130       & -2.0157 & -0.2032 & $0.6$& -51.11    &      $0.76$      & $-$         & $-$       &  $-$             & $-$              & $-$	      & $-$              	   &$-$ \\ 
131& -2.0164 & -0.2007 &$26.1$& -49.63& $1.39$     & $2.6\pm2.0$ & $+86\pm3$ &  $2.5^{+0.1}_{-1.0}$    & $10.6^{+0.4}_{-0.1}$    & $-$	      & $-$              	   &$\mathbf{63^{+3}_{-55}}$ \\ 
132& -2.0169 & -0.2007 &$27.1$& -48.37& $1.17$     & $2.6\pm0.7$ & $-87\pm8$ &  $2.4^{+0.1}_{-1.1}$    & $10.6^{+0.1}_{-0.1}$    & $-$	      & $-$              	   &$\mathbf{63^{+2}_{-42}}$ \\ 
133       & -2.0177 & -0.2009 & $0.6$& -51.53    &      $1.26$      & $-$         & $-$       &  $-$             & $-$              & $-$	      & $-$              	   &$-$ \\ 
134& -2.0178 & -0.2009& $7.6$& -49.00& $1.45$      & $5.0\pm1.5$ & $+86\pm9$ &  $2.4^{+0.1}_{-1.3}$    & $10.6^{+0.4}_{-4.5}$    & $-$	      & $-$              	   &$\mathbf{69^{+9}_{-39}}$ \\ 
135       & -2.0179 & -0.2005 & $0.8$& -45.00    &      $1.04$      & $-$         & $-$       &  $-$             & $-$              & $-$	      & $-$              	   &$-$ \\ 
136       & -2.0187 & -0.1970 & $0.2$& -41.00    &      $0.79$      & $-$         & $-$       &  $-$             & $-$              & $-$	      & $-$              	   &$-$ \\ 
137& -2.0449 & -0.1574& $7.2$& -45.21& $0.50$      & $3.3\pm0.5$ & $-24\pm1$ &  $1.0^{+0.5}_{-0.8}$    & $10.2^{+0.7}_{-1.9}$    & $-$	      & $-$              	   &$72^{+15}_{-6}$ \\ 
138       & -2.0645 &  0.2090 & $4.6$& -41.84    &      $0.58$      & $-$         & $-$       &  $-$             & $-$              & $-$	      & $-$              	   &$-$ \\ 
139       & -2.0706 &  0.1342 & $0.6$& -41.21    &      $0.61$      & $-$         & $-$       &  $-$             & $-$              & $-$	      & $-$              	   &$-$ \\ 
140       & -2.0724 &  0.1375 &$13.2$& -36.79    &      $0.89$      & $1.4\pm0.1$ & $+80\pm5$ &  $1.4^{+0.1}_{-0.2}$    & $9.5^{+0.2}_{-0.3}$     & $-$	      & $-$              	   &$69^{+12}_{-32 }$ \\ 
141       & -2.0853 & -0.2909 & $0.5$& -50.48    &      $1.83$      & $-$         & $-$       &  $-$             & $-$              & $-$	      & $-$              	   &$-$ \\ 
142       & -2.0873 & -0.3049 & $5.5$& -47.32    &      $0.62$      & $-$         & $-$       &  $-$             & $-$              & $-$	      & $-$              	   &$-$ \\ 
143       & -2.0875 & -0.3067 & $0.1$& -46.26    &      $0.95$      & $-$         & $-$       &  $-$             & $-$              & $-$	      & $-$              	   &$-$ \\ 
144       & -2.1038 &  0.2021 & $1.0$& -43.95    &      $0.96$      & $-$         & $-$       &  $-$             & $-$              & $-$	      & $-$              	   &$-$ \\ 
145       & -2.1099 &  0.1926 & $0.4$& -42.68    &      $0.59$      & $-$         & $-$       &  $-$             & $-$              & $-$	      & $-$              	   &$-$ \\ 
146       & -2.0152 & -0.2030 & $0.3$& -51.95    &      $1.28$      & $-$         & $-$       &  $-$             & $-$              & $-$	      & $-$              	   &$-$ \\ 
147       & -2.1208 & -0.2892 & $3.8$& -47.74    &      $0.84$      & $-$         & $-$       &  $-$             & $-$              & $-$	      & $-$              	   &$-$ \\ 
148       & -2.1228 & -0.2879 & $0.9$& -47.11    &      $0.66$      & $-$         & $-$       &  $-$             & $-$              & $-$	      & $-$              	   &$-$ \\ 
\label{features}
\end{longtable}
\end{longtab}